\documentclass[sigconf]{acmart}

\usepackage{flushend}

\usepackage{mathrsfs}

\usepackage{amsmath,amssymb,amsfonts}
\usepackage{graphicx}
\usepackage{textcomp}

\definecolor{mygreen}{HTML}{009000}
\definecolor{myred}{HTML}{C00000}
\definecolor{myyellow}{HTML}{DAA520}

\usepackage{booktabs}
\usepackage{multirow}
\usepackage{algorithm}
\usepackage[noend]{algpseudocode}

\usepackage{blindtext}
\usepackage{graphicx}
\usepackage{wrapfig}
\usepackage[caption=false]{subfig}
\usepackage[font=footnotesize]{caption}

\usepackage{tabularx}
\usepackage{amssymb}
\usepackage{soul}

\usepackage{enumitem}

\usepackage{pifont}

\usepackage[normalem]{ulem}

\usepackage{lipsum}  
\usepackage{diagbox}

\renewcommand\footnotetextcopyrightpermission[1]{} 

\AtBeginDocument{%
  \providecommand\BibTeX{{%
    \normalfont B\kern-0.5em{\scshape i\kern-0.25em b}\kern-0.8em\TeX}}}

\setcopyright{acmcopyright}
\copyrightyear{2020}
\acmYear{2020}
\acmPrice{15.00}
\acmDOI{10.1145/3400302.3415669}
\acmISBN{978-1-4503-8026-3/20/11}

\begin{document}

\title{NNgSAT: Neural Network guided SAT Attack \\ on Logic Locked Complex Structures}

\author{Kimia Zamiri Azar$^1$, Hadi Mardani Kamali$^1$, Houman Homayoun$^2$, Avesta Sasan$^1$}
\affiliation{\vspace{0.1cm}
\institution{$^1$ George Mason University, Fairfax, VA, USA.}
}
\email{{kzamiria, hmardani, asasan}@gmu.edu}
\affiliation{\vspace{0.1cm}
\institution{$^2$ University of California, Davis, Davis, CA, USA.}
}
\email{{hhomayoun}@ucdavis.edu}
\renewcommand{\shortauthors}{ }
\renewcommand{\shorttitle}{ }

\begin{abstract}

The globalization of the IC supply chain has raised many security threats, especially when untrusted parties are involved. This has created a demand for a dependable \emph{logic obfuscation} solution to combat these threats. Amongst a wide range of threats and countermeasures on logic obfuscation in the 2010s decade, the Boolean satisfiability (SAT) attack, or one of its derivatives, could break almost all state-of-the-art logic obfuscation countermeasures. However, in some cases, particularly when the logic locked circuits contain complex structures, such as big multipliers, large routing networks, or big tree structures, the logic locked circuit is \emph{hard-to-be-solved} for the SAT attack. Usage of these structures for obfuscation may lead a strong defense, as many SAT solvers fail to handle such complexity. However, in this paper, we propose a \textbf{neural-network-guided SAT attack} (\emph{NNgSAT}), in which we examine the capability and effectiveness of a message-passing neural network (MPNN) for solving these complex structures (SAT-hard instances). In NNgSAT, after being trained as a classifier to predict SAT/UNSAT on a SAT problem (NN serves as a SAT solver), the neural network is used to guide/help the actual SAT solver for finding the SAT assignment(s). By training NN on conjunctive normal forms (CNFs) corresponded to a dataset of logic locked circuits, as well as fine-tuning the confidence rate of the NN prediction, our experiments show that NNgSAT could solve 93.5\% of the logic locked circuits containing complex structures within a reasonable time, while the existing SAT attack cannot proceed the attack flow in them.

\end{abstract}


\settopmatter{printfolios=true}
\maketitle

\pagestyle{plain}


\section{Introduction}

The ever-increasing cost of integrated circuits (IC) manufacturing has pushed many semiconductor facilities to shift the integration procedure from vertical towards horizontal, where many stages of manufacturing require the involvement of third-party and offshore entities \cite{yeh2012trends}. Due to the lack of reliable monitoring over these entities, the trust in the IC supply chain has been called into question. Hence, as a result of the growth of globalization during IC manufacturing, numerous forms of hardware-based security threats have been emerged in the last two decades, including but not limited to overproduction, Trojan insertion, Reverse Engineering (RE), Intellectual Property (IP) theft, and counterfeiting \cite{tehranipoor2010survey, rostami2014primer, quadir2016survey, tehranipoor2017invasion}. 

To protect the IP/IC from being reverse engineered, overproduced, or stolen either in the manufacturing supply chain or in the field, researchers have studied various design-for-trust (DfTr) techniques, such as IC metering, hardware watermarking, split manufacturing, IC camouflaging, and logic locking \cite{alkabani2007active, kahng1998watermarking, roy2008epic, cocchi2013method, contreras2013secure, rajendran2012security, rajendran2013security}. Amongst these active/passive countermeasures, \emph{logic locking} a.k.a. \emph{logic obfuscation} \cite{roy2008epic, rajendran2012security}, as a proactive countermeasure against all these threats, hides the correct functionality of a chip by adding programmability into the design using post-fabrication programming values, referred to as the \emph{"key"}. The key values are initiated in and loaded from an on-chip tamper-proof non-volatile memory, and only once the correct key is provided, the chip behaves correctly. 

In the past decade, the concept and techniques used for logic obfuscation have gradually evolved, resulting in many forms and means of logic locking \cite{roy2008epic, baumgarten2010preventing, rajendran2012security, yasin2016sarlock, xie2016mitigating, yasin2017provably, shamsi2017cyclic, rezaei2018cyclic, azar2019coma, xie2017delay, shamsi2018cross, kolhe2019customlut, kamali2018lut, kamali2019full, kamali2020interlock}. At the same time, the security and strength of logic locking solutions have challenged by simultaneously evolving attack models \cite{subramanyan2015evaluating, el2015integrated, yasin2017removal, yasin2017security, shamsi2017appsat, shen2017double, xu2017novel, zhou2017cycsat, shen2019besat, shamsi2019icysat, sirone2020functional, chakraborty2018timingsat, azar2019smt, zamiri2019threats}. Amongst all existing attacks on logic locking, the Boolean satisfiability (SAT) based attack \cite{subramanyan2015evaluating, el2015integrated} has received the most attention. In the SAT attack, the adversary has access to (1) one successfully reverse-engineered yet locked netlist, and (2) the activated/functional IC (oracle). Also, in the activated/functional IC, the scan chain pins are available for any test/debug purposes. Hence, for each combinational logic part of the circuit that is accessible by the scan chain paths, the SAT attack could be applied independently. The SAT attack has an iterative structure, and in each iteration, a SAT solver is used to find a specific input, called discriminating input pattern (DIP) that finds two sets of keys producing different outputs. After finding all DIPs, the SAT solver can eliminate all incorrect keys leading to the extraction of the correct functionality of the circuit. 

The strength of the SAT solver comes from their Conflict-Driven Clause Learning (CDCL) ability. In each iteration of the SAT attack, a new SAT problem will be created, and the goal of the SAT solver is to find a satisfying (SAT) assignment for each SAT problem (per each iteration). The SAT problem is represented in conjunctive normal form (CNF) consisting of clauses, and each clause consists of one (a few) literal(s). The SAT solver tries to either assign or derive the value of each literal, and each assignment of value to a literal pushes the solver down into one of the branches of its decision tree. The decision tree of the SAT solver is built based on Davis–Putnam–Logemann–Loveland (DPLL) tree that is a complete backtracking-based search tree used for deciding the satisfiability of propositional logic formula. The traversal on the DPLL tree is based on a recursive DPLL algorithm leading to the finding of a SAT assignment (The DIP in each SAT attack iteration).

After the introduction of the SAT attack, many studies focus on proposing a logic locking countermeasure that resists against this powerful attack; However, almost all of them are vulnerable to a newer attack, and most of the newer attacks are derived from the SAT attack. For instance, since the SAT solver only works on directed acyclic graphs (DAG), a group of countermeasures evaluates the possibility of adding combinational cycles to the circuits \cite{rezaei2018cyclic, shamsi2017cyclic}. However, some studies introduce cyclic-based SAT attacks (cycSAT, beSAT, and icySAT \cite{zhou2017cycsat, shen2019besat, shamsi2019icysat}), in which a pre-processing graph-based re-encoding has been used before running the SAT attack to handle combinational cycles in the locked circuit. As another example, since the input format of the SAT solver is the conjunctive normal form (CNF), in some recent obfuscation techniques, the timing behavior (setup/hold time) of the circuit have been targeted that cannot be translated to CNF \cite{xie2017delay}. However, further investigation shows that this breed of obfuscation techniques is already broken using timingSAT and SMT attack \cite{chakraborty2018timingsat, azar2019smt}, where the delay is modeled using numerical constraints or a theory solver to de-obfuscate the locked circuit.

\subsection{Motivation and Contribution}

The SAT attack (or one of its derivatives) is able to break almost all existing logic locking solutions; However, when the locked circuit contains some complex structures, the SAT attack faces a challenging hardness, in which the locked circuit is a \emph{hard-to-be-solved} problem for the SAT solver. These complex structures could be any form of deep/symmetric/tree-based structures, particularly while they are built by the same basic Boolean logic. For instance, the work in \cite{subramanyan2015evaluating} has investigated that some locked circuits are challenging for the SAT solver in all contexts, such as circuits that contain \emph{multiplier}(s) (e.g. ISCAS-85 c6288). Similar to multipliers that have uniform distribution of \emph{AND}/\emph{XOR} trees for building bit-wise multiplication, ISCAS-85 c2670 contains \emph{AND-Tree} that makes it harder for the SAT solver. It gets worse when the size of these structures is large. For instance, in a typical \emph{AND}-Tree structure, for function $F=\bigwedge_{i=1}^N x_i$, that is encrypted using key values, each assignment to $k_1 . . . k_N$ in locked function $F^{locked}=\bigwedge_{i=1}^N (x_i \oplus k_i)$ is a singleton equivalence class, which makes it hard for the SAT solver to be solved when the size of the function is large.    
Furthermore, some recent studies show that depth/size/symmetry could affect the efficiency of the SAT solver significantly \cite{shamsi2018cross, kamali2019full}. This observation results in the introduction of a new category of obfuscation, called routing/interconnection obfuscation. In this breed of obfuscation solutions, key-programmable routing blocks (symmetric and deep MUX trees) have been used for obfuscation purposes. It is illustrated that when the routing block is in place, the depth of the DPLL tree in each iteration of the SAT solver would be increased drastically making the locked circuit a \emph{hard-to-be-solved} problem for the SAT solver. Accordingly, the \emph{runtime} of each iteration of the SAT attack would be increased significantly in routing-based obfuscation.

In this paper, to increase the strength of the existing SAT attack when complex structures are in place, we propose \emph{NNgSAT}, which is a neural network (NN) guided SAT attack on circuits that locked with or contain these SAT-hard structures. In NNgSAT, we get the benefit of a message-passing NN (MPNN) as a classifier. For cases where the SAT solver takes a long time to find the SAT assignment, the being trained neural network has been used to push/help/guide the SAT solver to find the assignment far sooner. Based on the iterative structure of the MPNN (iterative message passing), in NNgSAT, many of the complex structures will be solved leading us to successfully de-obfuscated the aforementioned SAT-hard instance. Given this goal, the main contributions of this paper are as follows:

\begin{enumerate}[leftmargin=*]
    \item We propose NNgSAT, a new attack on logic locking in which a neural network has been used and trained as a classifier to predict the satisfying assignment on a set of cases that the actual SAT solver cannot solve. 
    \item In the NNgSAT, we engage a 2-step training phase, in which we first exploit single bit supervision, as well as hyper-parameters as an initial training phase. This step is followed by the second step, where the training will be done on the obfuscated circuits obfuscated by small-size complex structures.
    \item We evaluate the impact of the confidence rate (CR) of the MPNN prediction on the effectiveness/performance of NNgSAT. 
    \item We propose a parallel SAT solving mechanism ensuring that the performance of NNgSAT could not be worse than the actual SAT in any form of obfuscations.
    \item Our experimental results show that, on average, NNgSAT successfully de-obuscates $\sim 93.5\%$ of cases that the original SAT attack fails to de-obfuscate them.
\end{enumerate}


\section{Background} \label{background}

Logic locking adds the possibility of post-fabrication programming into the supply chain; However, the introduction of the SAT attack raises a big question about the validity of this countermeasure. The SAT attack relies on some important assumptions: (1) The adversary has access to the successfully reverse-engineered netlist with full information about the location of the key gates, locking technique (algorithm), etc., (2) The adversary has access to an unlocked/activated chip, and the access to the scan chain pins are also available for any test/debug purposes. Relying on these assumptions, the SAT attack could be engaged independently on each combinational logic part, and many studies on logic locking show that this attack (or one of its derivatives) could break almost all of the existing logic locking techniques. 

\subsection{The SAT Attack: Iterative SAT solving of Miter Circuit}

\setlength{\textfloatsep}{5pt}
\begin{algorithm}[h]
\caption{SAT-based Attack Algorithm \cite{subramanyan2015evaluating}}
\label{SAT}
\begin{algorithmic}[1]

\small
\Function{SAT\_Attack}{Circuit~C$_{L}$, Circuit C$_{O}$}
    \State \emph{i} $\gets$ 0; 
    \State \emph{F$_{0}$} $\gets$ C$_{L}$(X, K$_{1}$, Y$_{1}$) $\land$ C$_{L}$(X, K$_{2}$, Y$_{2}$);
    \While {\emph{SAT}(\emph{F$_{i}$} $\land$ (Y$_{1}$ $\neq$ Y$_{2}$))} 
        \State X$_{d_i}$ $\gets$ sat\_assignment (F$_{i} \land $(Y$_1 \neq $Y$_2$)); 
        \State Y$_{d_i}$ $\gets$ C$_{O}$(X$_{d_i}$);
        \State \emph{F$_{i+1}$} $\gets$ \emph{F$_{i}$} $\land$ C$_{L}$(X$_{d_i}$, K$_{1}$, Y$_{d_i}$) $\land$ C$_{L}$(X$_{d_i}$, K$_{2}$, Y$_{d_i}$); 
        \State \emph{i} $\gets$ \emph{i+1};
    \EndWhile
    \State \emph{$K^*$}  $\gets$ sat\_assignment$_{K_1}$(\emph{F$_{i}$});
\EndFunction
\end{algorithmic}
\end{algorithm}

Getting inspired by the miter circuit used in the formal verification (equivalency check), in each SAT attack iteration, a miter circuit will be built to compare the (primary) outputs (PO) of two duplicated circuits ($F_0$ in line 3 of Alg. \ref{SAT}). Then, a SAT solver will be invoked to find a satisfying assignment for this miter circuit, in which for the same (primary) input (PI), noted as DIP, two different keys must produce two different outputs (line 5). Then, for finding the next DIP, the new pair of keys must produce the same output for all previous found DIPs (lines 6-7). Using this iterative structure, in each iteration, the SAT solver helps to find a new DIP, and when there is no new DIP, the miter and all constraints generated so far could identify the correct key (line 9). By using this flow, the SAT attack provides a very high convergence rate leading to breaking most of the existing logic locking solutions within a few minutes/hours.

\subsection{Hard SAT Instances in Logic Obfuscation}

Many logic locking techniques proposed in the recent few years trying to break the SAT attack using different approaches. For example, some techniques try to exponentially increase the number of iterations (DIPs) required to find the correct key \cite{yasin2016sarlock, xie2016mitigating, yasin2017provably}. Some techniques try to lock the timing of the circuit using either custom-designed cell or flip-flop relocation \cite{xie2017delay, zhang2018timingcamouflage}, which cannot be modeled by the SAT solver. In some other techniques, since the SAT attack is only applicable to DAG, the combinational cycles are added for obfuscation purposes, which might trap the iterative structure of the SAT attack in an infinite loop, or it leads to an incorrect key \cite{shamsi2017cyclic, rezaei2018cyclic, rezaei2019cycsat, roshanisefat2018srclock, roshanisefat2020sat}. Also, since the SAT solver is only applicable to combinational logic circuits, and the access to the scan chain is required for the SAT attack, some techniques try to block any unauthorized access to the scan chain \cite{wang2017secure, wang2018secure, guin2018robust, limaye2019robust, kamali2020scramble, kamali2020designing, roshanisefat2020dfssd} to resist against the SAT attack.   

Although these techniques were successful at the early stages, further investigation shows that many of them are vulnerable to either simple structural attacks or simple derivatives of the SAT attack in which the formulation is slightly changed or augmented by a pre-processor or co-processor to generate additional constraints that make the problem SAT solvable. Examples of such attacks are cyclic-based SAT attacks, structural-based integrated with SAT attack, unrolling-based SAT attacks, and SMT attack \cite{xu2017novel, el2017reverse, sirone2020functional, azar2019smt, zamiri2019threats}.

\begin{figure}
    \centering
    \includegraphics[width=\columnwidth]{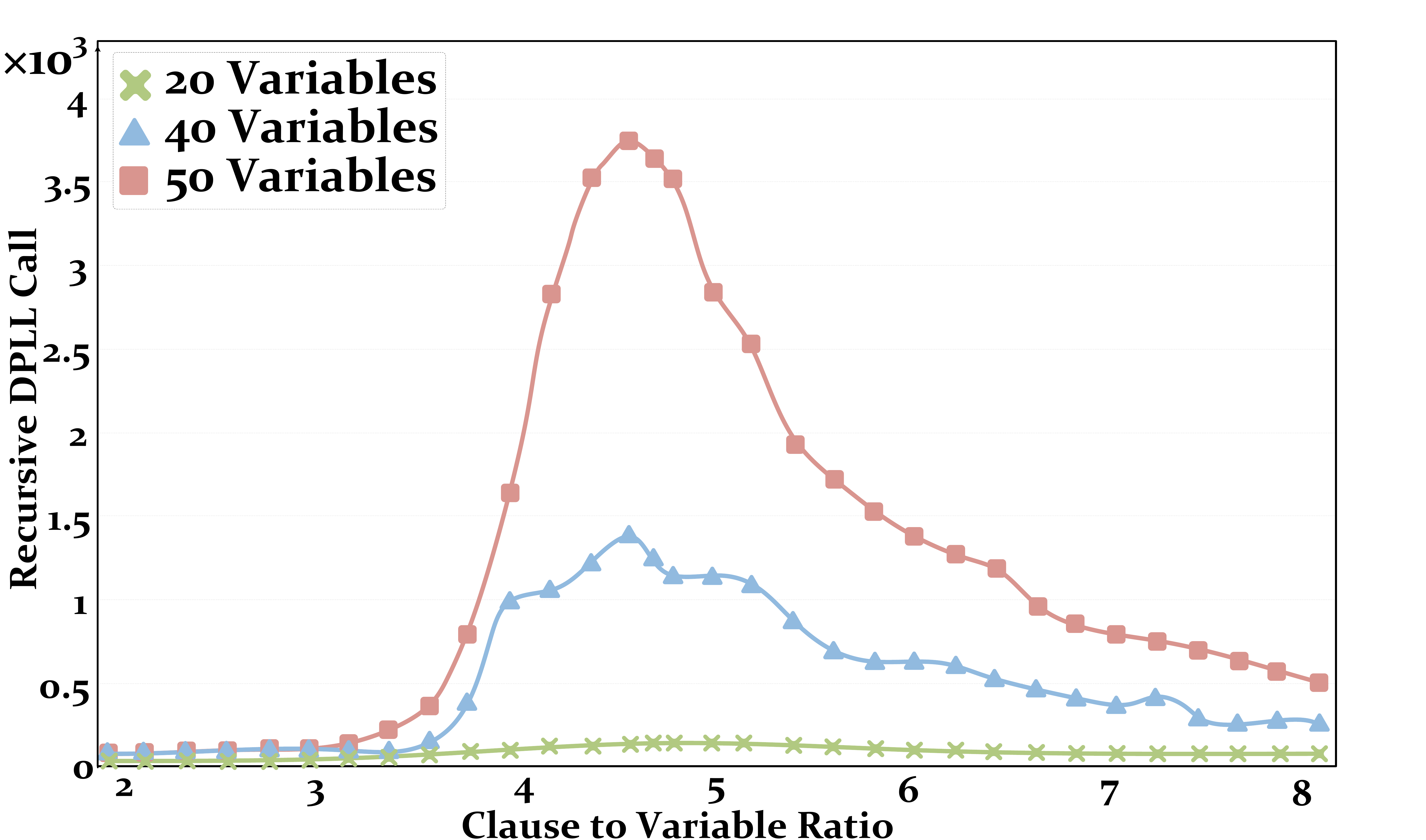}
    \vspace{-15pt}
    \caption{The Impact of Clause-to-Variable Ratio on the Median Number of Recursive DPLL Calls in a SAT Solver \cite{mitchell1992hard}.}
    \label{dpll_call}
\end{figure}

The SAT attack (or one of its derivatives) works perfectly fine in all aforementioned scenarios only while it does not face any form of \emph{hard-to-be-solved} instances. As a case of hard SAT instances, which brings difficulties for the SAT solvers, we could list deep or symmetric or tree-based structures, particularly while these structures are only built using the same basic gate, such as large multipliers, combinational systolic array modules, hierarchical routing blocks, big \emph{AND}-tree structures, etc. \cite{subramanyan2015evaluating, li2017provably, shamsi2018cross, kamali2019full}. Having such structures sends the corresponded CNF far away being \emph{under/over constrained}, and when the SAT problem is a medium-length CNF, it brings difficulties for the SAT solver. Fig. \ref{dpll_call} shows the number of recursive DPLL calls\footnote{DPLL algorithm is a recursive-based function that is the main part of the SAT solver (Line 5 in Algorithm \ref{SAT}) for finding the satisfying assignments.} for fixed-length 3-SAT CNFs, where the ratio of clauses to variables is varied from 2 to 8 \cite{mitchell1992hard, kamali2019full}. As demonstrated, when the SAT problem is a medium-length CNF (clauses to variables ratio from 4 to 6), it requires more DPLL calls than \emph{under/over constrained} CNFs (>6 or <4). This observation has been the motivation for techniques such as Cross-Lock and Full-Lock \cite{shamsi2018cross, kamali2019full}. In these techniques, the main block used for obfuscation is a key-programmable routing block, which helps to build an extremely large medium-length CNF with thousands of variables. Since it faces millions of DPLL calls for each iteration of the SAT attack, it exponentially increases the runtime of each iteration of the SAT attack. It is surprisingly clear that when a design has such \emph{hard-to-be-solved} instances, there is no chance for any of the existing attacks to break them, which motivates us to propose this new NN-based attack.

The question we aim to answer in this paper is whether we can train and use a neural network to predict a satisfying assignment to the literals of the SAT-circuit CNF and all constraints corresponded to each iteration of the SAT attack, particularly when the search tree is extremely deep. The goal is to save time and speed up the SAT attack. In cases the SAT solver cannot find the DIP for an iteration, we examine the effectiveness of a message-passing NN (MPNN) for predicting the DIP. This question is the motivation for the formulation of our proposed NNgSAT attack.

\section{Neural Network Learns the SAT Solving} \label{neural_back}

Glimpsing the progress of the SAT community starting around 1992 shows that before 2015 there were the most important conceptual advances resulting in the modern SAT algorithm based on CDCL. However, since 2015, the performance improvements has declined significantly \cite{oh2016improving}. On the other hand, the substantial ever-increasing the usage and application of the neural network (NN) on important problems raises a big question that \emph{"can a NN learn and improve the performance of the SAT solving?"}. More recently, a study proposes the first NN-based architecture that is designed for satisfiability problems, called \emph{NeuroSAT} \cite{selsam2018learning}. NeuroSAT is a message-passing neural network (MPNN) that learns to solve SAT problems after only being trained as a classifier to predict satisfiability. In NeuroSAT, it is shown that a trained NN on only toy problems could solve a bigger problem on its own. Also, based on the iterative structure of MPNN, more iterations at test time leads to solving bigger and even completely different domains than the problems it was trained on. For each problem, NeuroSAT starts guessing \emph{UNSAT} at the early stages with low confidence until it finds a solution, at which point it converges and return the satisfying assignment with very high confidence. The training phase of the NeuroSAT relies on a single bit of supervision, in which the only difference between a pair of SAT/UNSAT problem is one bit.

\begin{figure}
    \centering
    \vspace{-15pt}
    \subfloat[]{{\includegraphics[width=0.48\columnwidth]{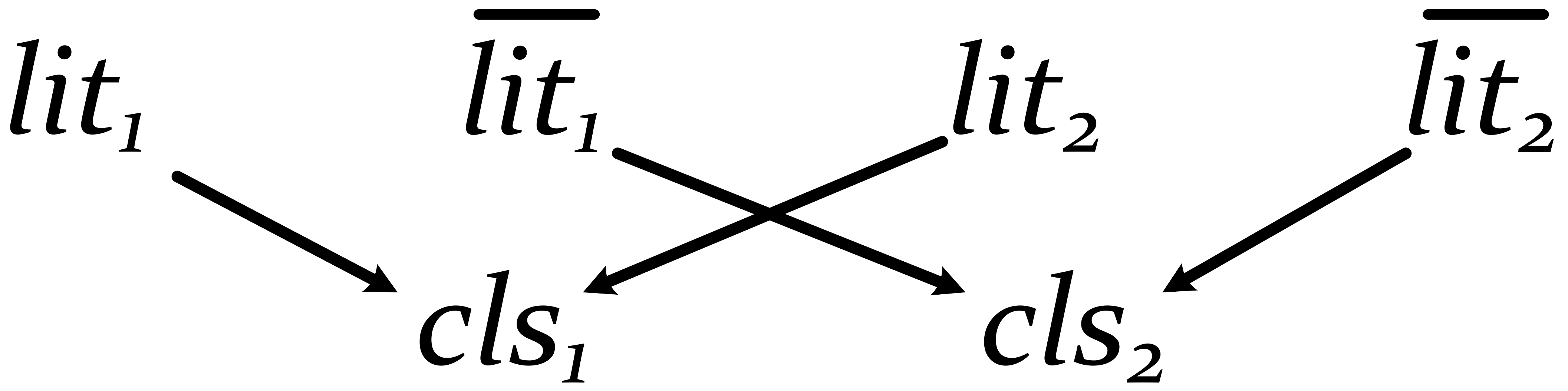}}} \hspace{2pt}
    \subfloat[]{{\includegraphics[width=0.48\columnwidth]{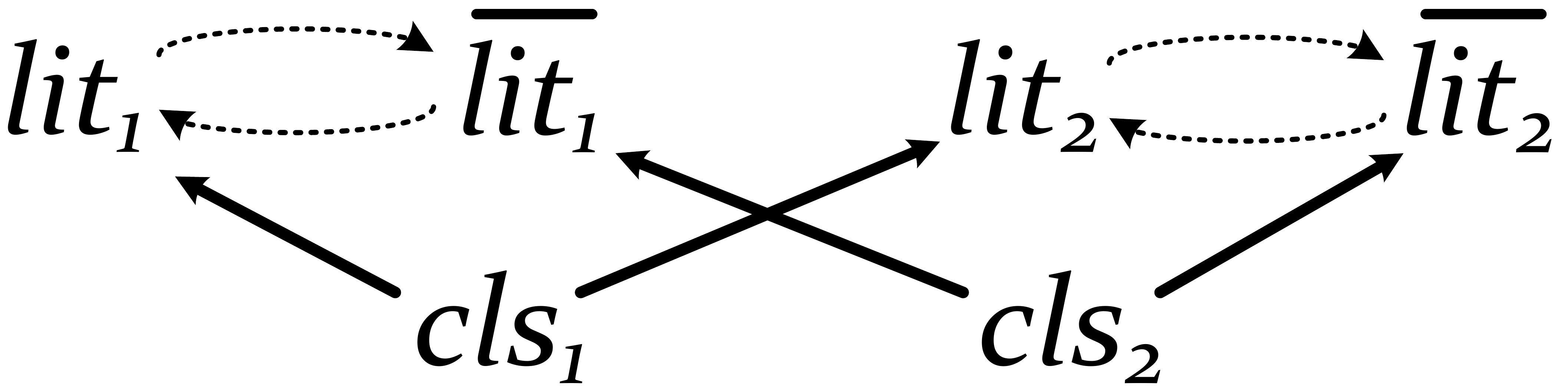}}} \hspace{2pt}
    \vspace{-10pt}
    \caption{High-level illustration of NeuroSAT operating on the graph representation of $(lit_1\vee lit_2) \wedge (\neg lit_1\vee \neg lit_2)$.}
    \label{graph}
\end{figure}

Since the SAT solver accepts the problems in CNF format, and since the SAT problem has a syntactic structure that could be encoded into a vector space, the best way to encode a SAT problem using a NN is to model it using an undirected graph. In this graph, there is one set of nodes each represents a literal, and one more set of nodes each represents a clause. For edges, there is a set of edges between each literal and each clause it appears in, and also there is another set of edges between each literal and its complement. Fig. \ref{graph}, shows a simple example how a SAT problem ($(lit_1\vee lit_2) \wedge (\neg lit_1\vee \neg lit_2)$) could be represented with an undirected graph with aforementioned rules. Nodes on the top represent each of the four literals, and nodes on the bottom represent each of the two clauses.

Assuming that this graph-based representation will be used for building the architecture of the NN, it could be parameterized by simply two vectors related to literals/clauses $(\textbf{L}_{init}, \textbf{C}_{init})$. Also, considering that a message-passing approach is used for the NN, the multi-layer perceptrons (MLP) could be $(\textbf{L}_{msg}, \textbf{C}_{msg}, \textbf{L}_{vote})$, and as a special form of recurrent NN (RNN), there exists two layer-norm long short-term memory (LSTM) networks for literals and clauses $(\textbf{L}_{u}, \textbf{C}_{u})$ \cite{hochreiter1997long}. Considering $n_v$ as the number of variables and $n_c$ as the number of clauses in a SAT problem, to model it to be solved using the iterative structure of the MPNN, at every time step $t$, the NN has a matrix $L^{(t)} \in  \mathbb{R}^{2n_v \times d}$ whose $i^{th}$ row contains the embedding\footnote{The row corresponding to a clause/literal referred to as the embedding of that clause/literal.} for the literal $\mathscr{l}_i$ and a matrix $C^{(t)} \in  \mathbb{R}^{n_c \times d}$ whose $j^{th}$ row contains the embedding for the clause $\mathscr{c}_j$, which are initialized with $\textbf{L}_{init}$ and $\textbf{C}_{init}$ respectively.

Using this formal definition, a single operation of voting (one iteration of message passing) consists of applying the following two updates, in which $M$ is the (bipartite) adjacency matrix defined by $M(i,j)= 1$ if $\{\mathscr{l}_i \in \mathscr{c}_j\}$, $Flip$ is the operator that takes a matrix $L$ and swaps each row of $L$ with the row corresponding to the literal’s negation, and both $L_h^{t} \in  \mathbb{R}^{2n_v \times d}$ and $C_h^{t} \in  \mathbb{R}^{n_c \times d}$ as the hidden states for $\textbf{L}_{u}$ and $\textbf{C}_{u}$ respectively, which initialized to zero matrices:

\begin{equation}
 (C^{t+1},C_h^{t+1}) \leftarrow \textbf{C}_u \big([C_h^{t},M^\top \textbf{L}_{msg}(L^{(t)})]\big) 
\end{equation}
\begin{equation}
(L^{t+1},L_h^{t+1}) \leftarrow \textbf{L}_u \big([L_h^{t}, Flip(L^{(t)},M\textbf{C}_{msg}(C^{(t+1)})]\big) 
\end{equation}

In this model, each iteration consists of two stages. First, each clause receives messages from its neighboring literals and updates its embedding based on the current embeddings  (Fig.\ref{graph}a). Next, each literal receives messages from its neighboring clause as well as from its complement and updates its embedding based on the current embeddings (Fig.\ref{graph}b). By using this scheme, after $T$ iterations, it could compute $L_*^{(T)} \leftarrow \textbf{L}_{vote}(L^{(t)}) \in \mathbb{R}^{2n_v}$, which is a single scalar representation of the vote for each literal, and then computes the average of the literal votes $y^{(T)}\leftarrow mean(L_*^{(T)}) \in \mathbb{R}$. This vote could be used as a parameter to show the confidence rate (CR) of the prediction provided by the NN. 

\begin{figure}
    \centering
    \vspace{-12pt}
    \includegraphics[width=\columnwidth]{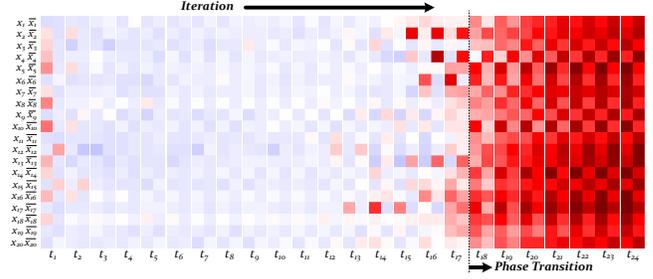}
    \vspace{-20pt}
    \caption{The Sequence of Literal Votes (Confidence Rate (CR)) $L_*^{(1)}$ to $L_*^{(24)}$ in the MPNN Running on a SAT Problem Containing 20 Variables ($x_{1..20}$) \cite{selsam2018learning}.}
    \label{vote}
\end{figure}

Fig. \ref{vote} provides a better representation to understand how iterative-based MPNN could be used to predict and guess the satisfying assignment for a SAT problem with a scalar representation of the vote (CR). It illustrates the sequence of literal votes for 24 iterations $L_*^{(1)}$ to $L_*^{(24)}$, \big($L_*^{(T)} \leftarrow \textbf{L}_{vote}(L^{(t)})$\big), as the NN runs on a SAT problem. As shown in Fig. \ref{vote}, to clarify the voting of each literal, $L_*^{(T)}$ is reshaped to be an $\mathbb{R}^{n_v \times 2}$ matrix so that each literal is paired with its complement (the $i^{th}$ row contains the scalar votes for $x_i$ and $\bar{x_i}$). As shown in Fig. \ref{vote}, for iterations at the early stages, almost every literal is voting \emph{UNSAT} with low confidence (light blue). Then, a few scattered literals start voting \emph{SAT} for the next few iterations. However, it is not dominant to affect the mean vote. In most cases with a sudden change, there is a phase transition, and all the literals (and hence the network as a whole) start to vote \emph{SAT} with very high confidence (dark red). After this phase transition, the vote for each literal converges, and the network does not need to continue evolving. 

\begin{figure}
    \centering
    \vspace{-15pt}
    \subfloat[]{{\includegraphics[width=0.23\columnwidth]{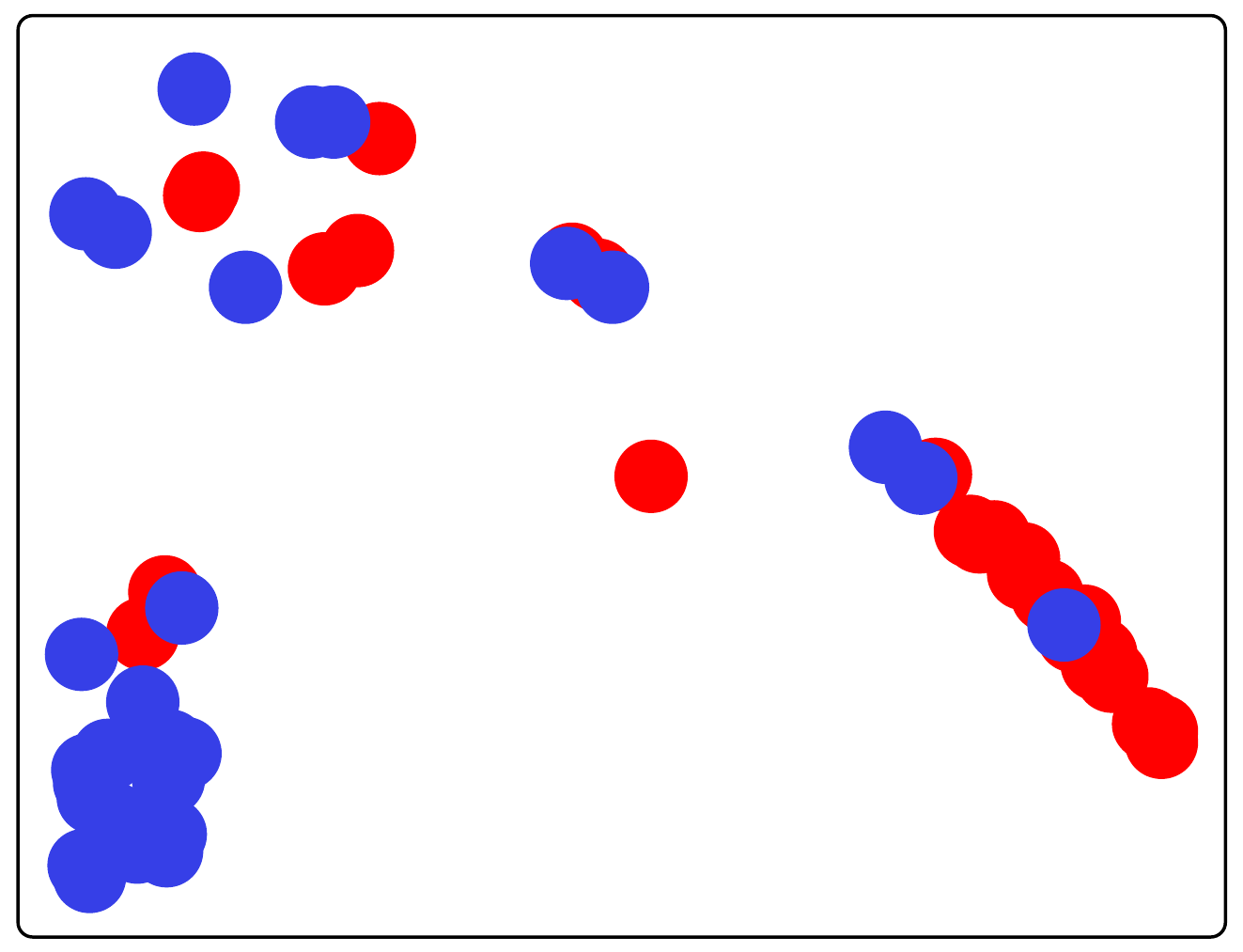}}} \hspace{2pt}
    \subfloat[]{{\includegraphics[width=0.23\columnwidth]{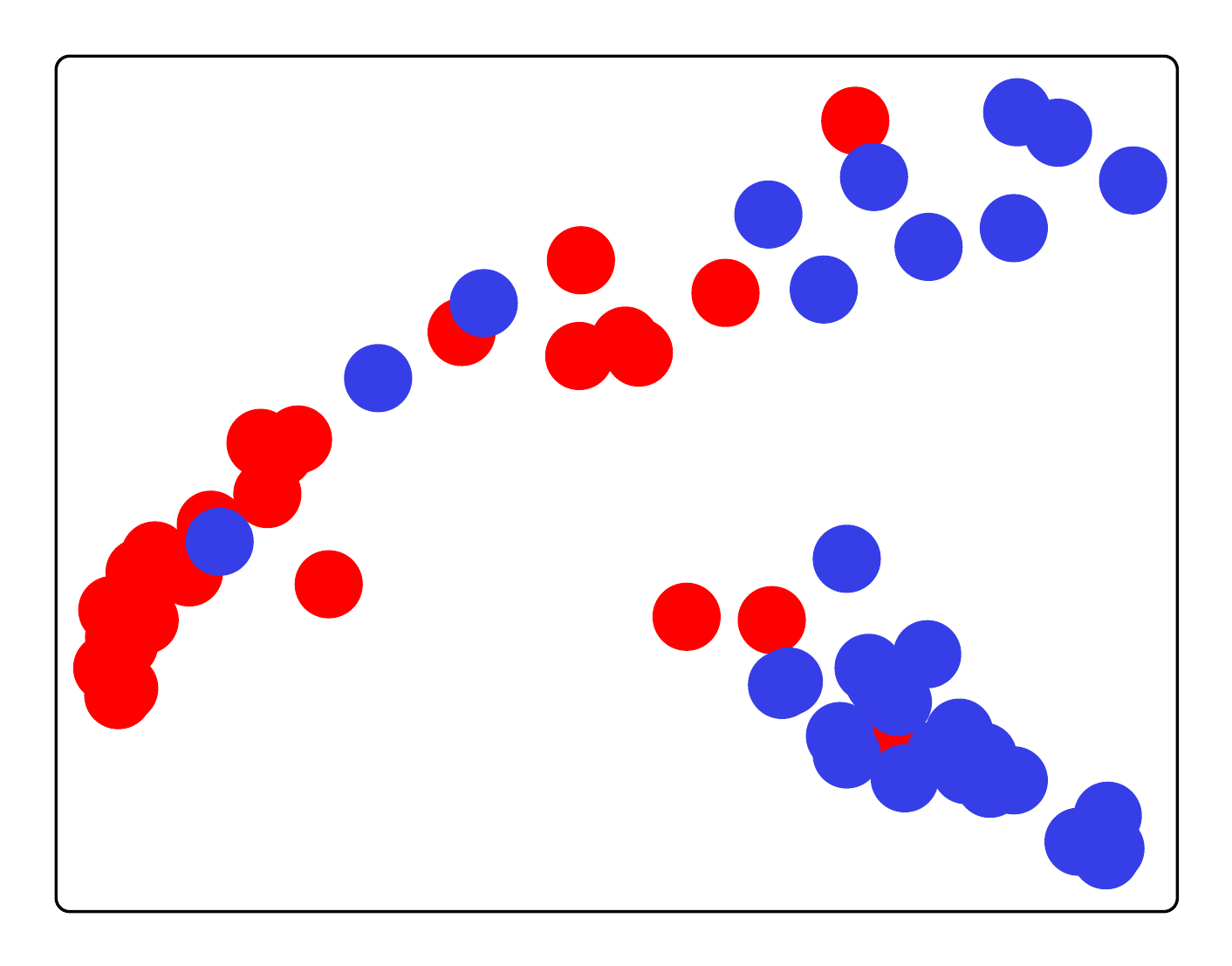}}} \hspace{2pt}
    \subfloat[]{{\includegraphics[width=0.23\columnwidth]{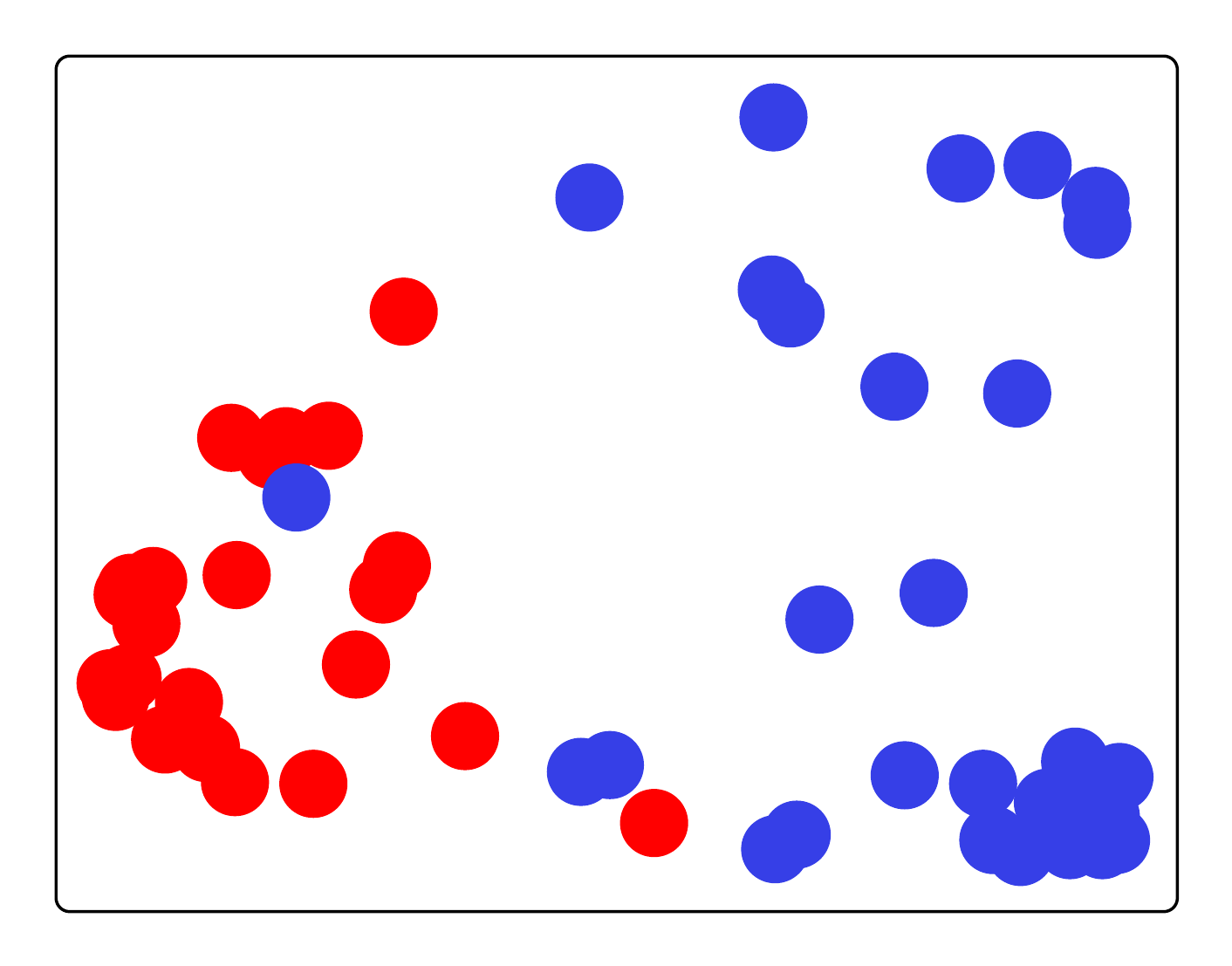}}} \hspace{2pt}
    \subfloat[]{{\includegraphics[width=0.23\columnwidth]{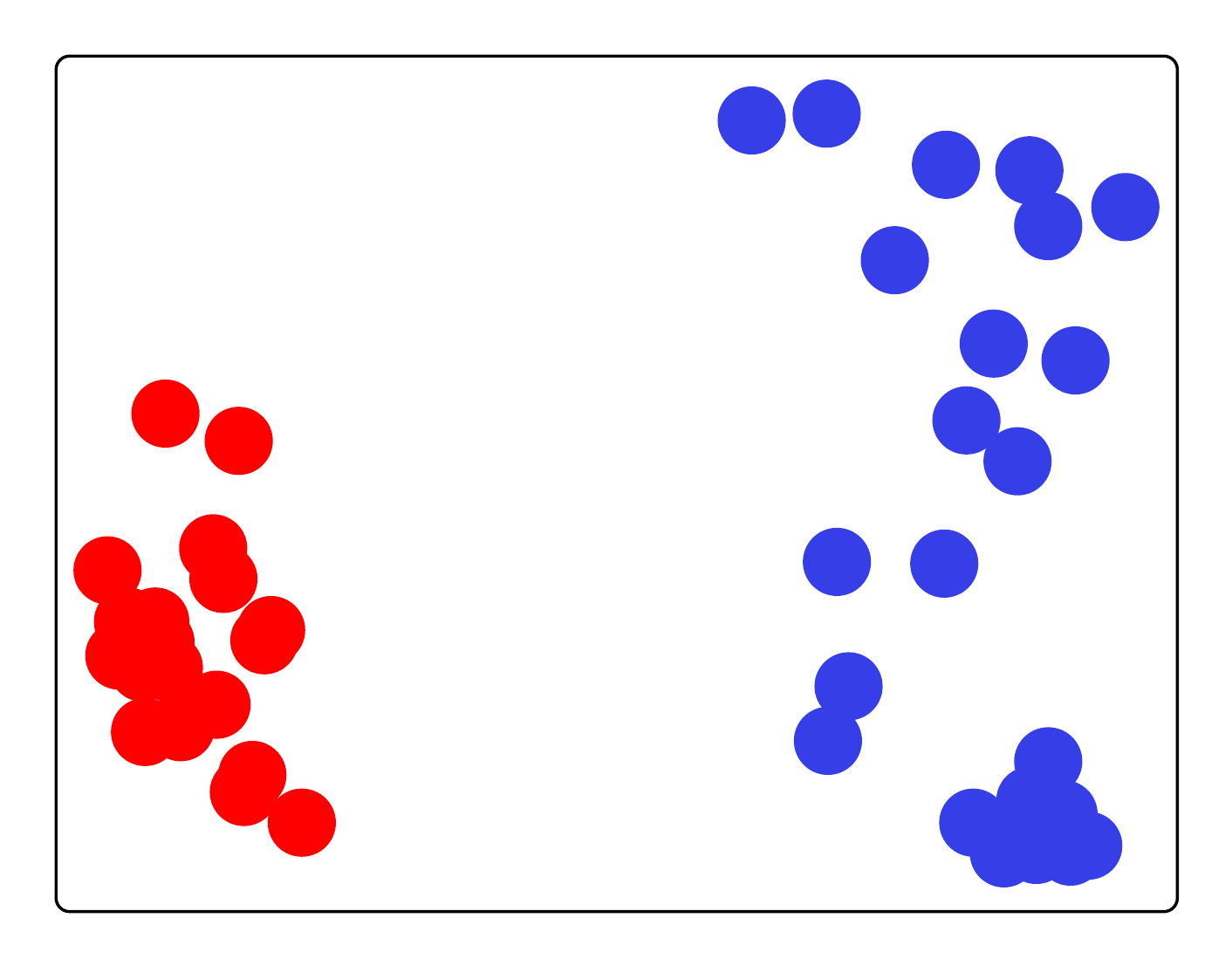}}}
    \vspace{-10pt}
    \caption{Extracting the Satisfying Assignment using 2-Clustering K-means.}
    \label{kmeans}
\end{figure}

As shown in Fig. \ref{vote}, most of the variables have one literal vote distinctly darker (higher confidence rate) than the other. Also, the dark votes have all approximately the same color tone, and the same color tone could be seen for light votes. Based on this distinguishable coloring, it turns out that there exists a meaningful relationship between the satisfying assignment and these patterns (Darker votes are \emph{1}s and Lighter ones are \emph{0}s). However, to have a more reliable decoding solution, it could be translated using a 2-clustering mechanism calculated by the k-means algorithm. As shown in Fig. \ref{kmeans}, applying 2-clustering on the literal votes in each iteration ($L^t$) helps to distinguish between 0s and 1s to correctly extract the satisfying assignment (blue and red dots denote literals set to 0 and 1, respectively).

\begin{figure}[t]
    \centering
    \vspace{-15pt}
    \includegraphics[width=\columnwidth]{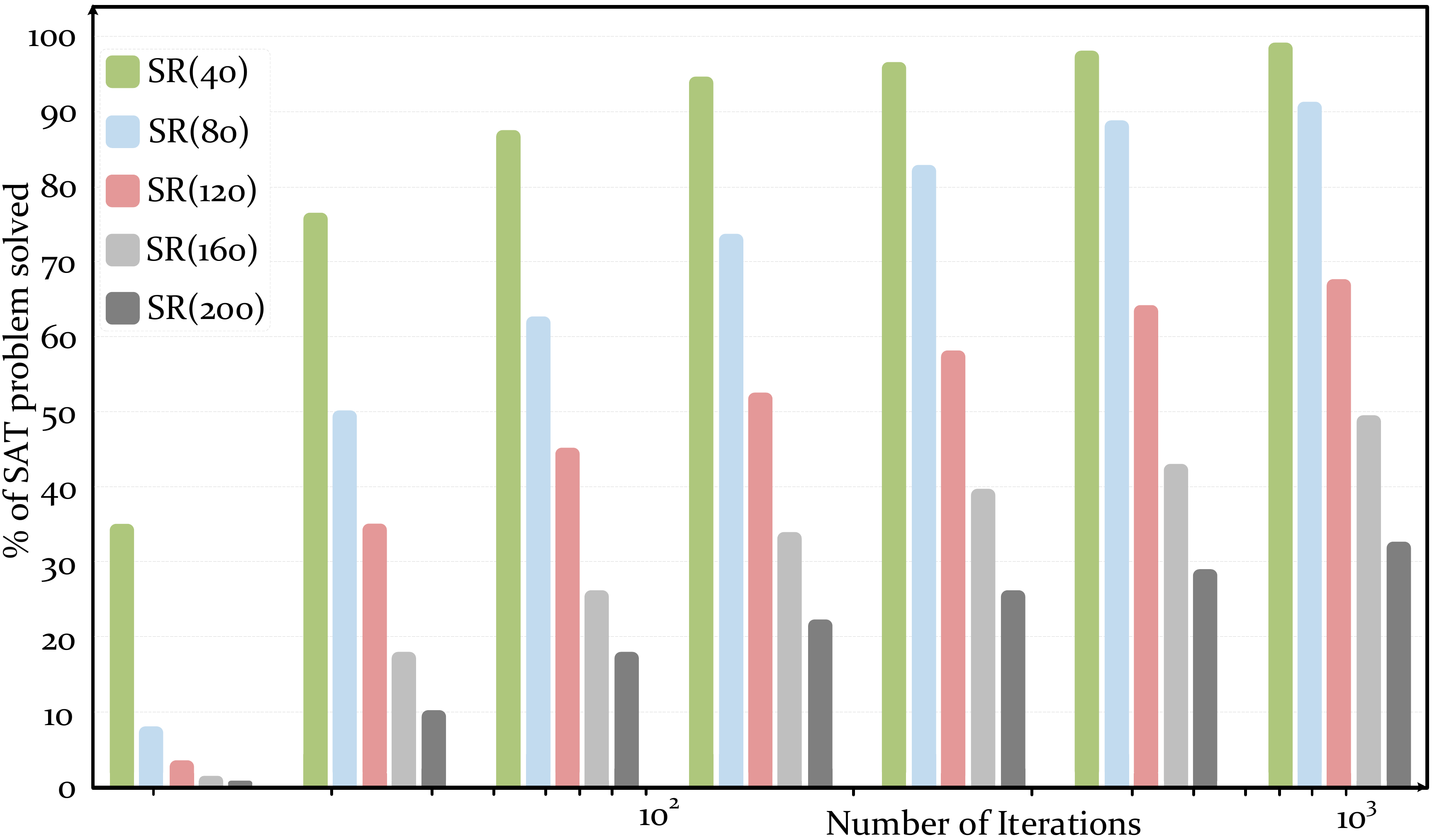}
    \vspace{-18pt}
    \caption{NeuroSAT’s success rate on $SR(n)$ for a range of $n$ as a function of the number of iterations $T$.}
    \label{successrate}
\end{figure}

A key observation of the NN usage for solving SAT problems is that it can solve SAT problems that are far larger than the models used during training. This is when the NN runs more iterations of message passing leading to find the satisfying assignment. As an example, Fig. \ref{successrate} shows the success rate of the NN on $\textbf{SR}(n)$\footnote{$\textbf{SR}(n)$ is a distribution over pairs of random SAT problems on $n$ variables, in which one element of the pair is satisfiable, the other is unsatisfiable, and the two differ by negating only a single literal occurrence in a single clause.} for a range of $n$ as a function of the number of iterations $T$. The NN is only trained on $SR(U(10, 40))$ (the number of variables $n$ is uniformly sampled from between 10 and 40 during training), however, as shown in Fig. \ref{successrate}, even though it is trained on $SR(40)$ and below, it solves SAT problems sampled from $SR(n)$ for $n$ much larger than $40$ by simply running for more iterations.

The observation that NeuroSAT can solve problems that are substantially larger and more difficult than it ever saw during training (by simply running for more iterations), motivates us to engage an MPNN to examine the effectiveness of this form of the solver for de-obfuscation purpose particularly while there exists \emph{hard-to-be-solved} structures in the locked circuit.

\section{NNgSAT:A NN guides The SAT Attack} \label{proposed}

\begin{figure}[t]
    \centering
    \vspace{-15pt}
    \includegraphics[width=\columnwidth]{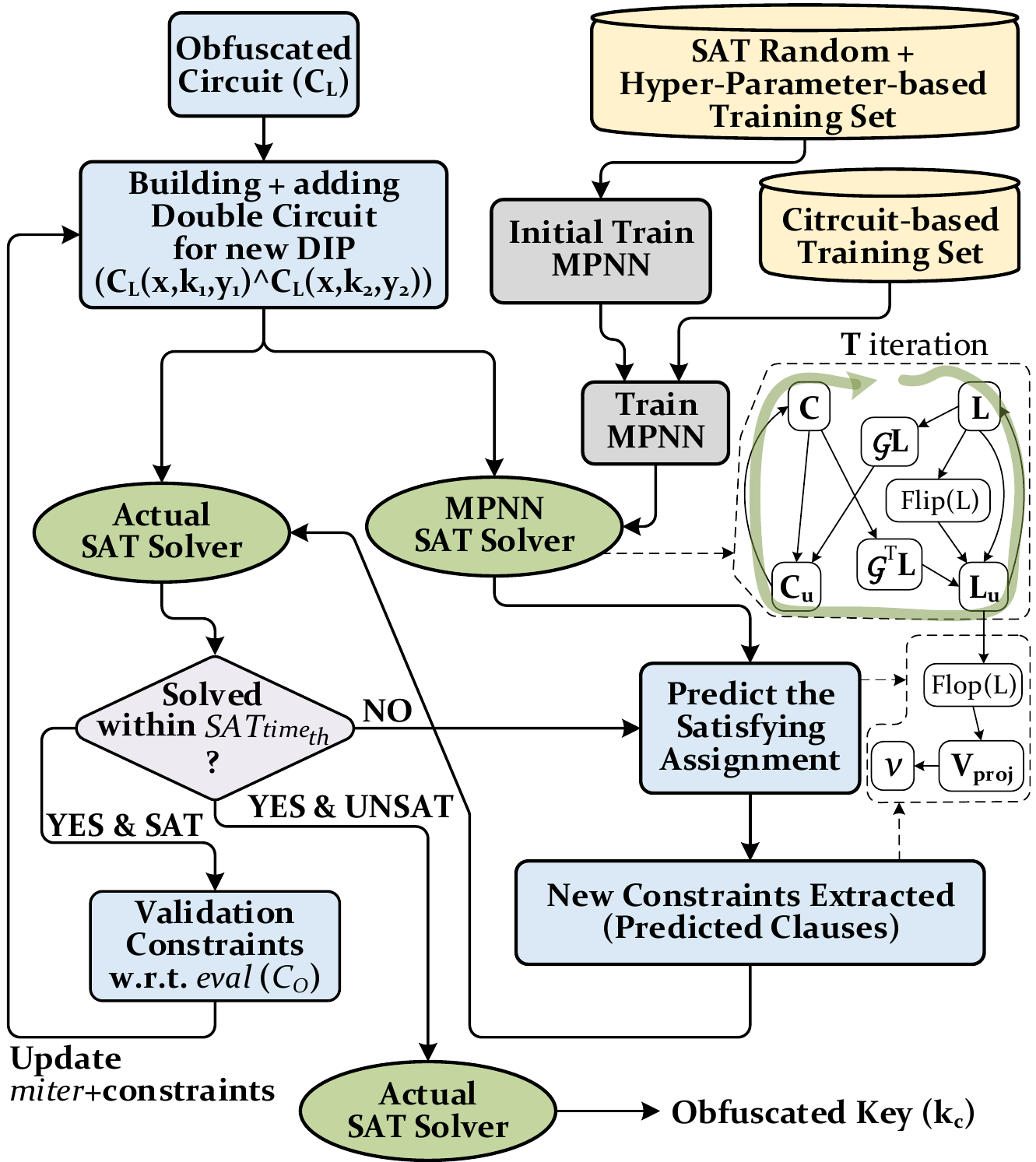}
    \vspace{-15pt}
    \caption{The Major Steps of NNgSAT Attack.}
    \label{nngsat}
\end{figure}

As its name implies, in NNgSAT, getting inspired from NeuroSAT, a \emph{message passing neural network (MPNN)} has been engaged and trained on the specific SAT problems obtained from logic locked circuit to be used as a guide for the SAT solver within the SAT attack. Given the SAT attack algorithm illustrated in Algorithm \ref{SAT}, to get the benefit of the MPNN, after being trained using the generated data set(s), the MPNN-based SAT solver is called in parallel with the actual SAT solver per each SAT iteration. Fig. \ref{nngsat} provides an overview of the major steps in the NNgSAT attack. As can be seen, most steps are similar to the traditional SAT attack. However, per each iteration, after updating the CNF of $miter$+constraints, and adding the new double-circuit for finding the new DIP, both the SAT solver and MPNN-based SAT solver will be called in parallel to solve the updated CNF. Based on a pre-defined threshold time ($SAT{time_{th}}$), if the actual SAT solver could find the satisfying assignment before $SAT{time_{th}}$, the MPNN-based SAT solver will be skipped, and for the next step both solvers will be called again. However, in those cases that the SAT solver could not find the satisfying assignment within $SAT{time_{th}}$, a part of the predicted satisfying assignment by the MPNN-based SAT solver, which have the highest literal votes (CR), will be extracted as a new (guiding) learned constraint, to help the actual SAT solver for finding the precise satisfying assignment.

Since the MPNN-based SAT solver is called in parallel with the actual SAT solver, after $SAT{time_{th}}$, we assume that $T$ iterations of the MPNN-based SAT solver is executed ($T$ times of message passing). As shown in Fig. \ref{nngsat}, in MPNN-based SAT solver, for a CNF with $n_c$ clauses and $n_v$ variables, set of clauses ($C$) and set of literals ($L$) will be initialized with $n_c$ clauses and $2n_v$ variables ($n_v$ variables + $n_v$ negated variables). Then, similar to the NeuroSAT, for $T$ iterations of message passing, $C$ and $L$ would be updated in two stages: (1) clause updating: based on the current embeddings of the literals it contains ($\forall \mathscr{c}, \mathscr{c} \leftarrow C_u (\mathscr{c}, \sum_{\mathscr{l} \in \mathscr{c}} L_{msg(\mathscr{l})}$), (2) literal updating: based on the current embeddings of the clauses it occurs in, plus the current embedding of its negation ($\forall \mathscr{l}, \mathscr{l} \leftarrow L_u (\mathscr{l}, \sum_{\mathscr{l} \in \mathscr{c}} C_{msg}(\mathscr{c}), \bar{\mathscr{l}}$). Also, to have better convergence, a $n_c\times 2n_v$ sparse matrix $\mathcal{G}$ has been used, in which $\mathcal{G}_{i,j}$=1 if and only if the $i^{th}$ clause contains the $j^{th}$ literal. Similar to NeuroSAT, the $Flip(L)$ function swaps the first half of the rows of a matrix ($L$) with the second half. Also, to acquire better projection, after $T$ iterations, the NN flops $L$ to proceed the prediction. The $Flop(L)$ function concatenates the first half of the rows of a matrix with the second half along the second axis. Then, the flopped $L$ followed by a projection ($V_{proj}$ as an MLP, to project into an $n_v$-dimensional vector for completing the prediction (scalar vote)).

In NNgSAT, the number of message passing operations (the MPNN iterations) depends on the value of $SAT{time_{th}}$ and scalar vote (CR). We assumed that when we reach at time=$SAT{time_{th}}$, $T$ iterations have been done in the MPNN. However, it might be possible that after $T$ iterations of the MPNN, the prediction is \emph{UNSAT} or even \emph{SAT} with low CR. Hence, to avoid misguiding the SAT attack, more iterations are also required in MPNN-based SAT solver. Hence, to not lose the chance of solving by the actual SAT attack and to continue using the MPNN, we do not stop any of the solvers (either actual or MPNN) after $SAT{time_{th}}$. Instead, we define a few thresholds for the scalar vote (CR) in the MPNN. When we reach each threshold, we extract the prediction provided by the MPNN, and use those variables that have the highest scalar vote (CR) as a guided constraint. Then, we run a new actual SAT solver learned by the \emph{miter}+constraints extracted from all previous iterations (successfully done), as well as guided by the MPNN. Hence, based on the pre-defined confidence thresholds, a few instances of the actual SAT solver would be run in parallel after $SAT{time_{th}}$, each is guided by different constraints predicted by the MPNN with different CR. Amongst all actual SAT solvers executing in parallel, the first actual SAT instance that returns \emph{SAT}, will be used as the solution for the current SAT iteration, all others will be skipped.

\begin{figure}[t]
    \centering
    \subfloat[]{{\includegraphics[width=\columnwidth]{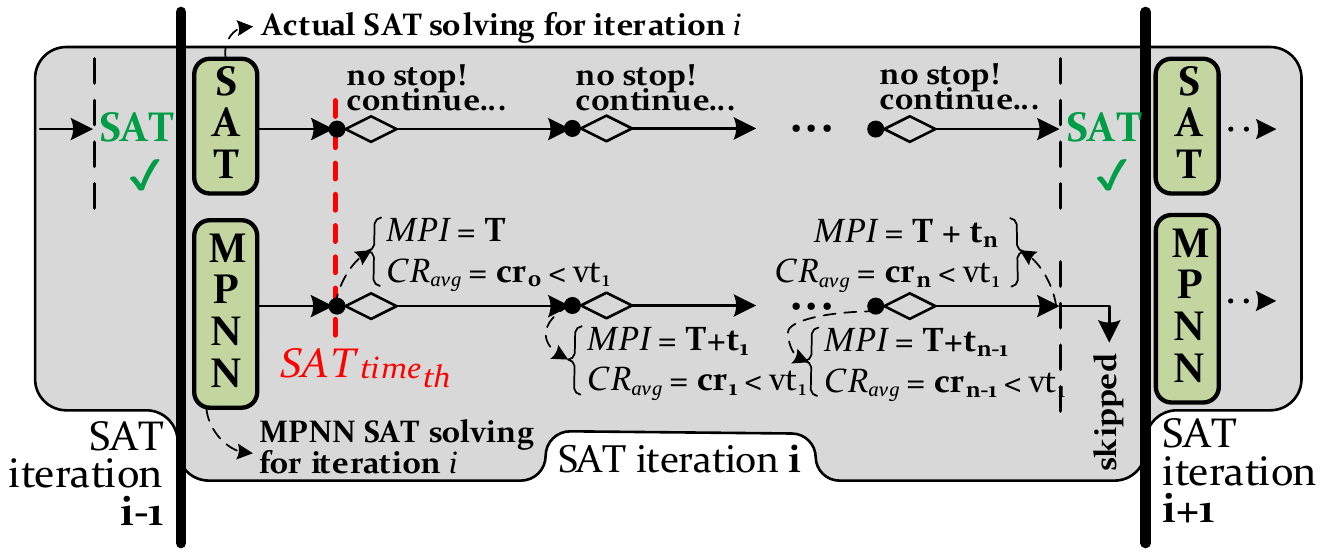}}} \\ 
    \subfloat[]{{\includegraphics[width=\columnwidth]{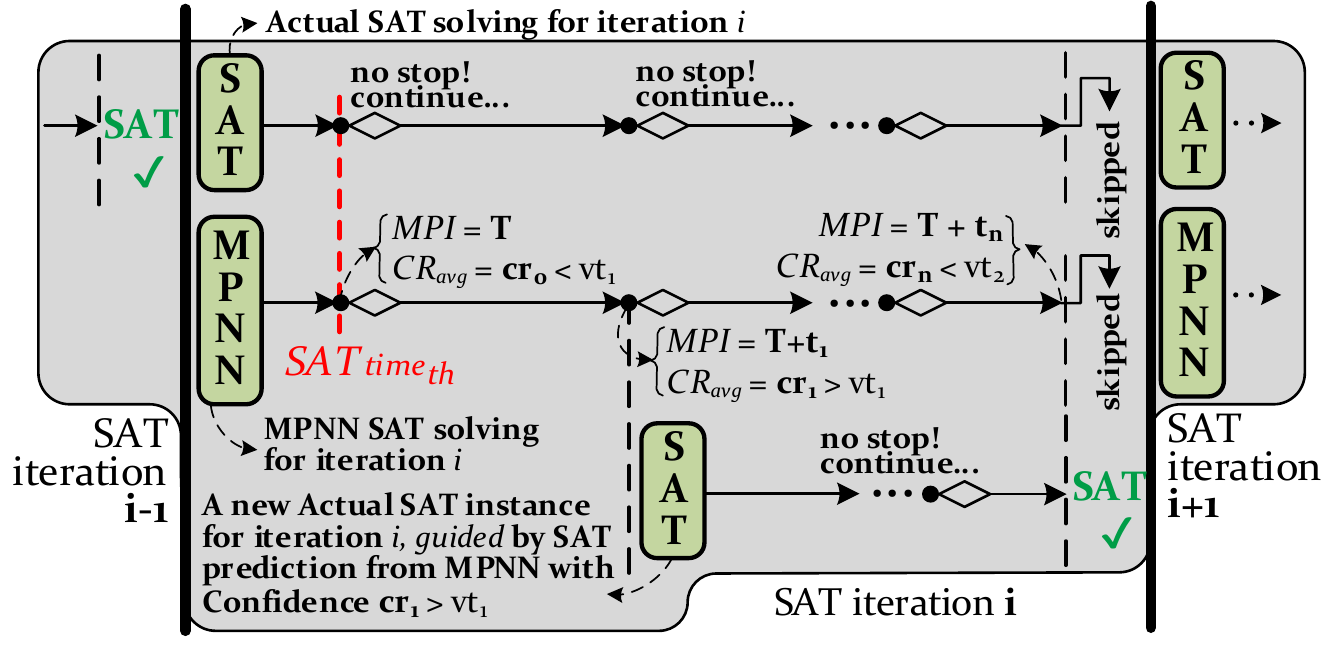}}} \\ 
    \subfloat[]{{\includegraphics[width=\columnwidth]{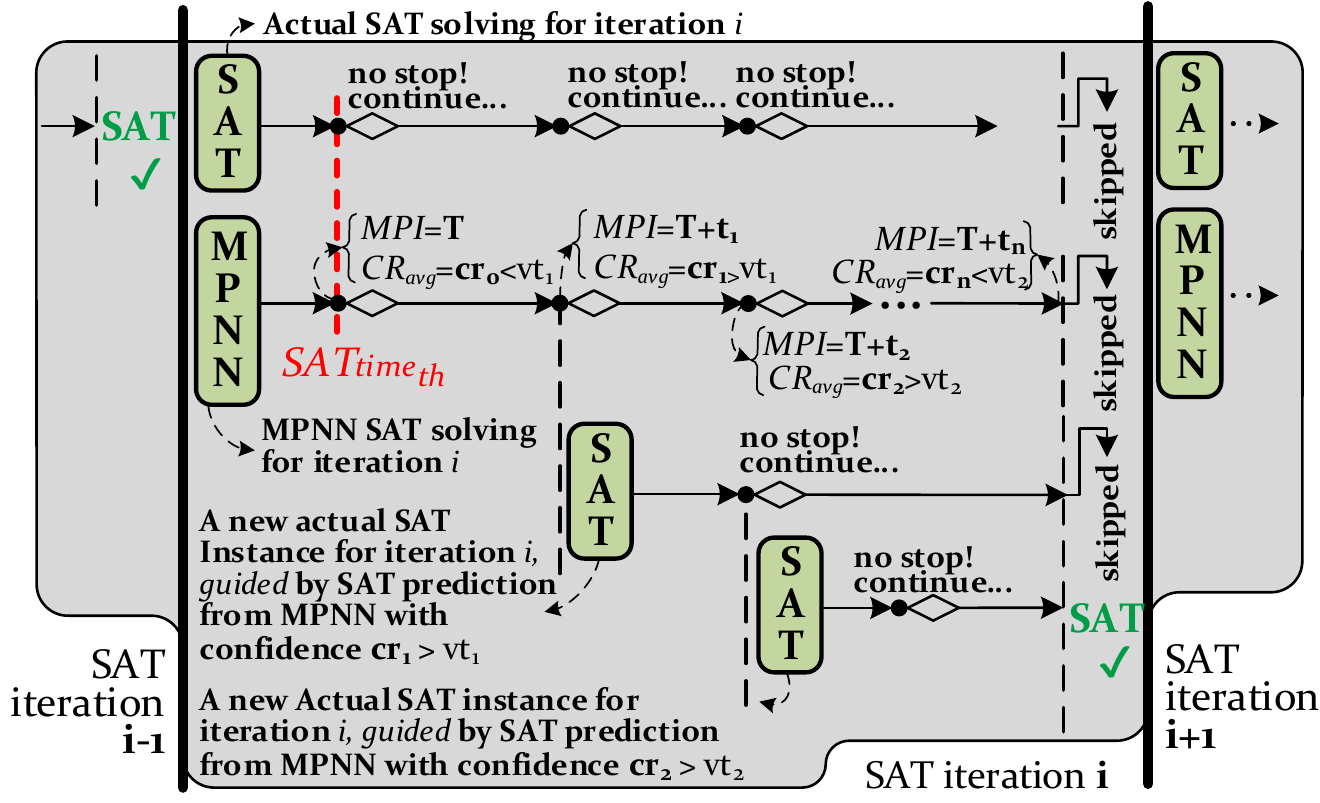}}} 
    \vspace{-10pt}
    \caption{Parallel Running of Actual SAT Guided by the MPNN with Different Scalar Votes (CRs).}
    \label{parallel_sat}
\end{figure}

Fig. \ref{parallel_sat} shows three different scenarios in NNgSAT w.r.t. the parallel SATs. In Fig. \ref{parallel_sat}(a), after $SAT{time_{th}}$, the CR is less than the first (minimum) CR threshold ($vt_1$). So, we continue running both to see which solver reaches the next state. In this case, even before reaching the first CR threshold ($vt_1$), the actual SAT was able to find the result. So, we skip the MPNN, and the SAT attack goes to the next SAT iteration. In Fig. \ref{parallel_sat}(b), after $SAT{time_{th}}$, the CR is still less than the first (minimum) CR threshold ($vt_1$). However, before finding the SAT assignment by the main actual SAT solver, the MPNN reaches the first CR threshold ($vt_1$). When the MPNN reaches a CR threshold, it generates a prediction as a SAT assignment, and variables with the highest CR (Those variables with CR$_{bit}$>90\%) will be extracted as a set of the learned clause. Then a new actual SAT solver (the second instance) will be started guided by the extracted learned clause acquired by the MPNN. Now, two actual SAT solvers and MPNN execute simultaneously, and in this case, after a while, the second SAT instance that was guided by the MPNN were able to find the SAT assignment far sooner, and two other instances (the first actual SAT solver and MPNN) will be skipped. Similar scenarios are demonstrated in Fig. \ref{parallel_sat}(c), where the third SAT solver returns faster.  

It should be noted that in the MPNN-based SAT solver, there is no dependency between the learned parameters and the size of the SAT problem. However, since the SAT attack iteratively adds a new double circuit, as well as a lot of learned clauses found in the previous iteration, the size of the SAT problem, would be increased extremely, which makes it very memory-intensive. Hence, to query the MPNN with more scalability, building the sparse matrix $\mathcal{G}$ would be based on a limited number of non-eliminated variables and clauses. Although we would like to pass all clauses/variables to the MPNN, in many cases, the problem would not fit into the available memory. Hence, we traverse the learned clauses in ascending size order, collecting part of them that does not exceed a fixed cutoff. Also, we only query the MPNN on random subsets of the clauses for problems that exceed the cutoff. 

Since both SAT/UNSAT will happen in each SAT attack\footnote{Each iteration of the SAT attack returns SAT meaning that a new DIP is found. When there exists no more DIP, it returns UNSAT, revealing the correct key with one more SAT solving.}, the MPNN should be able to predict UNSAT as well. To help construct proofs for UNSAT problems, the MPNN that has been trained on a large dataset, in which every unsatisfiable problem contains a small contradiction, learns to detect these contradictions instead of searching for satisfying assignments. Then, the variables involved in the contradiction can be extracted and enable constructing a resolution proof more efficiently. The NN cannot predict that a SAT problem is UNSAT with high confidence, and it almost never guesses SAT on an UNSAT problem. It only predicts SAT, once it has found one satisfying assignment. Hence, it could be considered as a certificate of satisfiability.

\begin{table*}[ht]
\footnotesize
\centering
\vspace{-15pt}
\caption{Specifications of the Modified (\_m) Benchmark Circuits (cxxxx\_m/bxx\_m) After the Insertion of Complex Structures. \vspace{-8pt}}
\label{benchs_modified}
\setlength\tabcolsep{2pt} 
\setlength\extrarowheight{1pt}
\begin{tabular}{@{} l||ccc|c|c|c|c @{}}
\toprule
Circuit & \# of Inputs & \# of Outputs & \# of Gates & \# of Multipliers & \# of Crossbars & \# of LUTs & \# of AND-trees \\
\cmidrule(r){1-1} \cmidrule(r){2-4} \cmidrule(r){5-5} \cmidrule(r){6-6} \cmidrule(r){7-7} \cmidrule(r){8-8}
c2670\_m & 287 & 161 & 5,793 & 1 of 12$\times$14, 1 of 18$\times$18 & 1 of 24$\times$20 & 4$\times$LUT-6, 2$\times$LUT-8 & 2$\times$AND-16 \\
c3540\_m & 87 & 47 & 7,605 & 2 of 7$\times$10, 1 of 20$\times$16, 1 of 18$\times$22 & 1 of 18$\times$15, 1 of 20$\times$16 & 4$\times$LUT-5, 3$\times$LUT-8 & 2$\times$AND-15, 1$\times$AND-26 \\
c5315\_m & 206 & 152 & 10,851 & 3 of 11$\times$14, 1 of 12$\times$22 & 2 of 19$\times$14, 1 of 36$\times$30 & 6$\times$LUT-4, 2$\times$LUT-6, 2$\times$LUT-7 & 1$\times$AND-32 \\
c6288\_m$^*$ & 85 & 74 & 7,378 & 1 of 20$\times$25 & 3 of 9$\times$9, 1 of 12$\times$15 & 2$\times$LUT-8, 1$\times$LUT-9, 2$\times$LUT-12 & 4$\times$AND-22, 1$\times$AND-52 \\
c7552\_m & 224 & 121 & 12,722 & 1 of 12$\times$18, 1 of 22$\times$26 & 2 of 15$\times$15, 1 of 30$\times$36 & 3$\times$LUT-6, 1$\times$LUT-15 & 2$\times$AND-15, 3$\times$AND-27 \\
b14\_m & 289 & 308 & 20,407 & 2 of 8$\times$12, 2 of 12$\times$14, 1 of 36$\times$34 & 2 of 26$\times$22 & 2$\times$LUT-11 & 2$\times$AND-16, 1$\times$AND-44 \\
b15\_m & 488 & 526 & 19,348 & 1 of 29$\times$25 & 1 of 30$\times$32, 1 of 36$\times$36 & 2$\times$LUT-11, 1$\times$LUT-16 & 3$\times$AND-33, 1$\times$AND-36 \\
b17\_m & 1452 & 1512 & 47,972 & 2 of 17$\times$21, 1 of 18$\times$22 & 1 of 26$\times$22, 1 of 32$\times$30 & 4$\times$LUT-6, 2$\times$LUT-12 & 2$\times$AND-42 \\
b20\_m & 527 & 516 & 38,856 & 1 of 8$\times$12, 1 of 32$\times$26 & 1 of 24$\times$26, 1 of 60$\times$64 & 2$\times$LUT-11, 1$\times$LUT-14 & 1$\times$AND-12, 1$\times$AND-48 \\
b22\_m & 767 & 757 & 35,658 & 1 of 12$\times$14, 2 of 11$\times$11, 1 of 14$\times$18 & 2 of 17$\times$14, 1 of 24$\times$22 & 4$\times$LUT-5 & 1$\times$AND-16, 4$\times$AND-29  \\

\bottomrule
\multicolumn{5}{l}{$^*$: c6288 contains 16$\times$16 multiplier by itself.}
\end{tabular}
\vspace{-10pt}
\end{table*}

\subsection{Training Dataset Generation} \label{training_gen}

With extensive analysis, we observe that our proposed NN-based SAT attack works best if the NN is trained in two phases. First, the NN needs to be initially trained using simple and small random SAT problems with single-bit supervision to learn how distinguishing between SAT and UNSAT. Also, with a minor change compared to the NeuroSAT, as illustrated in Fig. \ref{nngsat}, the network is not only fine-tuned using single-bit supervision, but it also followed by training with a single set of hyper-parameters as a coarse heuristic that broadly assigns higher CR to selected variables. This coarse learning also helps us to have more reliable (with higher confidence) predictions for UNSAT problems \cite{selsam2019guiding}. 

Then as the second phase of the training, the NN is trained specifically for a set of only small-size circuits that contain small-size of the complex structures, listed as follows:

\begin{enumerate}[leftmargin=*]
    \item $n \times m$ bit-wise multiplier built by \emph{AND/XOR} trees ($m,n \leq $ 8).
    \item $n \times m$ crossbar network built by $2$-to-$1$ MUXes ($m,n \leq $ 16). 
    \item $n-$input look-up-tables (LUT) built by $2$-to-$1$ MUXes ($n \leq $ 8). 
    \item $n$-to-$1$ \emph{AND}-tree structures, built by \emph{AND2} (binary-tree).
\end{enumerate}

We insert these structures into the circuit as a part of the original circuit, as well as a part of the obfuscation module (key-programmable crossbars or LUTs). The sizes that are selected for this phase of the training must be small enough to make them solvable by the traditional SAT attack. Also, to maximize what is learned by the network from these special structures, we engage circuits as simple as possible. To do that, we add these structures into a circuit that only contains wires connecting outputs to the inputs. In fact, this dataset only consists of these structures. 

\section{NNgSAT Evaluation} \label{results}

To investigate the performance and effectiveness of NNgSAT, we selected and used large circuits from ISCAS-85 and ITC-99 benchmark suite, as summarized in Table \ref{benchs}. For the neural network, we used \emph{NeuroSAT} with minor changes in training and message passing calculations described in Section \ref{proposed}. To compare the results of NNgSAT with the actual SAT attack, we use the traditional SAT attack \cite{subramanyan2015evaluating}. We also use MiniSAT as the SAT solver of both traditional SAT and NNgSAT. The timeout has been set to 2$\times$10$^5$ seconds ($\sim$ two days). All experiments have been done on a Dell PowerEdge R620 equipped with 28-core Intel Xeon E5-2670 2.50GHz and 64GB of RAM. 

For the training phase, we built two different training datasets. First, as initial training, to force the network to learn problems substantive, we define a distribution \textbf{SR}(\emph{n}) over pairs of random SAT problems, where one element is SAT, and the other is UNSAT, and they differ by negating only a single literal occurrence in a single clause. After fine-tuning over these pairs, we also trained a single set of hyper-parameters as a coarse heuristic that broadly assigns higher CR to selected variables, and training is done using ADAM optimizer \cite{kingma2014adam} with a constant learning rate of 10$^{-4}$. Second, we built 20,000 obfuscated circuit samples (based on the small sizes from Section \ref{training_gen}), each obfuscated using key-programmable LUTs/crossbars and solved by the actual SAT solver. Then, we train the NN based on the output of the actual SAT solver per each iteration.

To show the effectiveness of NNgSAT on complex structures, we modify the selected benchmark circuits by embedding some large-size complex structures into the design. To test the efficacy of the NNgSAT in finding satisfiable assignment in real-circuits, we then embedded larger-than-trained structures into the raw benchmark circuits as follows:

\begin{enumerate}[leftmargin=*]
    \item $n \times m$ bitwise multipliers built by \emph{AND/XOR} trees (8<$m,n$<32).
    \item $n \times m$ crossbar network built by $2$-to-$1$ MUXes (16<$m,n$<36). 
    \item $n-$input look-up-tables (LUT) built by $2$-to-$1$ MUXes ($n$<16). 
    \item $n$-to-$1$ \emph{AND}-tree structures, built by \emph{AND2} (binary-tree).
\end{enumerate}

\begin{table}[t]
\footnotesize
\centering
\caption{Specifications of the Raw Benchmark Circuits \vspace{-8pt}}
\label{benchs}
\setlength\tabcolsep{1.5pt} 
\begin{tabular}{@{} l cccccccccc @{}}
\toprule
Circuit: & c2670 & c3540 & c5315 & c6288$^*$ & c7552 & b14 & b15 & b17 & b20 & b22 \\
\cmidrule(r){1-1} \cmidrule(r){2-11}
\cmidrule(r){1-1} \cmidrule(r){2-11}
\# of Inputs    & 233 & 50 & 178 & 32 & 207 & 277 & 485 & 1452 & 522 & 767\\
\cmidrule(r){1-1} \cmidrule(r){2-11}
\# of Outputs    & 140 & 22 & 123 & 32 & 108 & 299 & 519 & 1512 & 512 & 757 \\
\cmidrule(r){1-1} \cmidrule(r){2-11}
\# of Gates     & 1,193 & 1,669 & 2,307 & 2,416 & 3,513 & 9,014 & 8,367 & 36,770 & 19,682 & 29,162 \\
\bottomrule
\multicolumn{5}{l}{$^*$: c6288 is a 16$\times$16 multiplier by itself.}
\end{tabular}
\end{table}

By using these structures, we built new and modified benchmark circuits to be evaluated using NNgSAT, as described in Table \ref{benchs_modified}. As shown, for each design, 1,2, or 3 multipliers with different sizes, 1, or 2 crossbars, few LUTs, and \emph{AND}-tree(s) have been inserted to be considered as a part of the original design. It is worth mentioning that in a few cases, particularly for smaller circuits, since the wiring is limited, we had to add extra primary inputs/outputs into the design to provide the required nets. A comparison between the original circuits listed in Table \ref{benchs} and modified (\_m) ones listed in Table \ref{benchs_modified} shows that some of the modified circuits have more PI/PO. Furthermore, these modules are embedded with the highest conservation to avoid emerging any form of incompatibility/instability, such as avoiding the creation of nonsense logic function, bypass logic, and combinational cycles. It should be noted that the size of these structures must be large enough to be considered as \emph{hard-to-be-solved} instances. On the other hand, the size must be small enough to keep the overhead less than the acceptable overhead threshold (e.g. \%5-\%10)\footnote{Although these structures incur up to 5$\times$ overhead in the benchmark circuits, in the real applications, such as well-known microprocessors, the overhead of these structures would be less than \%5.}.

\begin{table*}[ht]
\footnotesize
\centering
\vspace{-20pt}
\caption{Average Execution Time Comparison between NNgSAT and the Traditional SAT Attack \cite{subramanyan2015evaluating}. \vspace{-10pt}}
\label{sat_exe}
\setlength\tabcolsep{1.8pt} 
\setlength\extrarowheight{1pt}
\begin{tabular}{@{} l||cc|cc|cc|cc|cc|cc|cc|cc @{}}
\toprule
Circuit: & \multicolumn{2}{c}{c2670\_m} & \multicolumn{2}{c}{c3450\_m} & \multicolumn{2}{c}{c6288\_m} & \multicolumn{2}{c}{c7552\_m} & \multicolumn{2}{c}{b15\_m} & \multicolumn{2}{c}{b17\_m} & \multicolumn{2}{c}{b20\_m} & \multicolumn{2}{c}{b22\_m} \\
\cmidrule(r){1-1} \cmidrule(lr){2-3} \cmidrule(lr){4-5} \cmidrule(lr){6-7} \cmidrule(lr){8-9} \cmidrule(lr){10-11} \cmidrule(lr){12-13} \cmidrule(lr){14-15} \cmidrule(lr){16-17}
Attack: & NNgSAT & SAT & NNgSAT & SAT & NNgSAT & SAT & NNgSAT & SAT & NNgSAT & SAT & NNgSAT & SAT & NNgSAT & SAT & NNgSAT & SAT \\
\cmidrule(r){1-1} \cmidrule(lr){2-3} \cmidrule(lr){4-5} \cmidrule(lr){6-7} \cmidrule(lr){8-9} \cmidrule(lr){10-11} \cmidrule(lr){12-13} \cmidrule(lr){14-15} \cmidrule(lr){16-17}

1C18$\times$16, 4LUT10 & 1607.8 & 3858.6 & 1205.7 & 1205.7 & 3055.7 & timeout & 2490.5 & 2490.5 & 4661.8 & 4661.8 & 2207.4 & timeout & 2764.9 & 11706.6 & 7294.7 & timeout \\
2C18$\times$16, 2LUT12 & 7207.4 &  {timeout} &  {4184.7} &  {timeout} &  {3112.9} &  {timeout} &  {8679.2} &  {timeout} &  {13212.9} &  {timeout} &  {10766.2} &  {timeout} &  {12647.1} &  {timeout} &  {7554.6} &  {timeout} \\
1C24$\times$18, 4LUT10 & 2494.7 & 12029.1 & 4284.5 & 4284.5 &  {1285.1} &  {timeout} & 2570.7 & 16372.8 & 7206.4 & 12473.8 &  {5622.1} &  {timeout} &  {9223.7} &  {timeout} & timeout & timeout \\

2C24$\times$18, 2LUT12		   & 5670.8 & timeout & 5991.6 & 5991.6 & 2976.1 & timeout & 7708.4 & timeout & 11382.4 & timeout & 10684.5 & timeout & 10700.5 & timeout & 14382.8 & timeout \\
		                       
1C36$\times$30, 4LUT10		   & 6850.7 & 23074.5 & 5573.9 & timeout & 2403.3 & timeout & 6189.5 & 21688.8 & 7681.1 & timeout & 9034.8 & timeout & 9927.8 & timeout & 11042.1 & timeout \\
		                       
2C36$\times$30, 2LUT12		   & 5034.7 & timeout & 6408.3 & timeout & 3606.6 & timeout & 11684.8 & timeout & 4896.4 & timeout & 7221.5 & timeout & 10082.8 & timeout & 11894.7 & timeout \\

2C18$\times$16, 2C24$\times$18 & 7920.7 & timeout & 6270.8 & 18604.4 & 1967.2 & timeout & 5608.5 & 12685.8 & 6672.3 & timeout & 4956.7 & timeout & 5501.2 & timeout & 5974.6 & timeout \\

2C18$\times$16, 2C36$\times$30 & 8134.8 & timeout & 6303.1 & timeout & 3007.2 & timeout & 6122.7 & timeout & 9501.8 & timeout & 6079.3 & timeout & 6442.6 & timeout & timeout & timeout \\

2C24$\times$18, 2C36$\times$30 & 8507.3 & timeout & 7190.8 & timeout & 4622.9 & timeout & 8642.8 & timeout & timeout & timeout & 12820.3 & timeout & timeout & timeout & 10941.5 & timeout \\
\bottomrule
\multicolumn{17}{l}{$^*$ timeout = 2$\times$10$^5$ Seconds.\textcolor{white}{...................} $^+$ $n_c$C$m\!\times\!n$ = $n_c$ crossbars with size $m \times n$ controlled by the key. \textcolor{white}{...................} $^+$ $n_l$LUT$n$ = $n_l$ LUTs with $n$ inputs configured with the key.}
\end{tabular}
\vspace{-10pt}
\end{table*}

After building the modified benchmark circuits, one more step is the obfuscation to be assessed by NNgSAT. For obfuscating the benchmark circuits, we also used the key-programmable form of these structures. We insert large-size LUTs\footnote{For LUTs, the configuration bits are assumed as the key inputs.} and key-programmable crossbars with different sizes. More precisely, crossbars with size 18$\times$16, 24$\times$18, and 36$\times$30, as well as LUTs with 10 and 12 inputs are added into the design.  

Table \ref{sat_exe} illustrates the average execution time of the traditional SAT attack \cite{subramanyan2015evaluating} and our proposed NNgSAT on obfuscated benchmark circuits. As shown, in many cases, the original SAT attack fails to extract the satisfying assignment(s) within the allowed time threshold when a large-size of such complex structures are in place. However, after being trained over single-bit supervision and hyper-parameters and expanding the training over only small-size complex structures, NNgSAT could break almost all obfuscated circuits within a few hours. Also, NNgSAT is affected far less by the size of the circuit or the size of the obfuscation module, showing that correctly being trained, the network could also improve itself for especially large circuits that never saw during the training. Besides, unlike accelerated SAT solutions (e.g. parallelism of SAT) that linearly improve the performance, in almost all cases, the NNgSAT improvement is super-linear.   

\begin{figure}[t]
    \centering
    \vspace{-10pt}
    \subfloat[]{{\includegraphics[width=0.5\columnwidth]{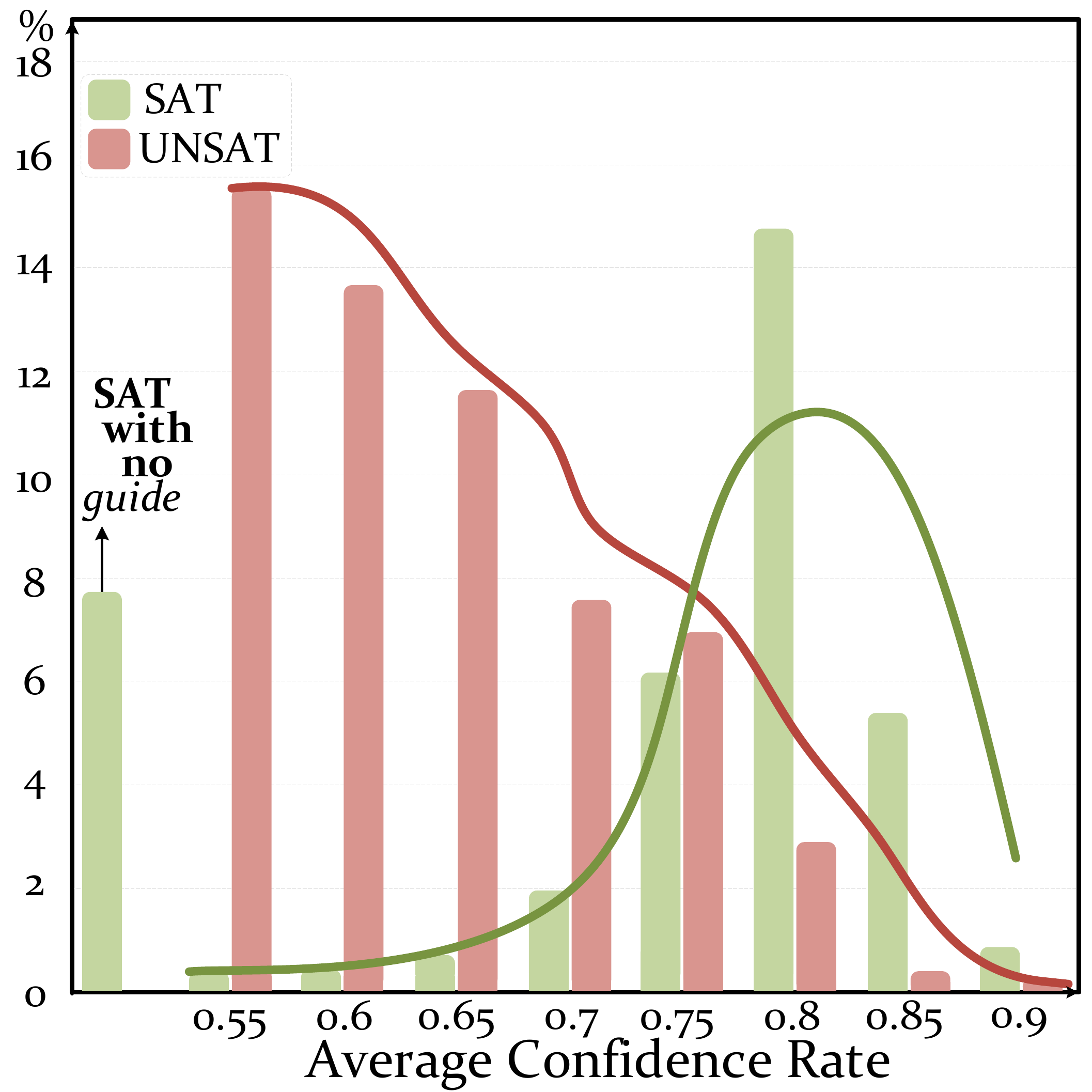}}}
    \subfloat[]{{\includegraphics[width=0.5\columnwidth]{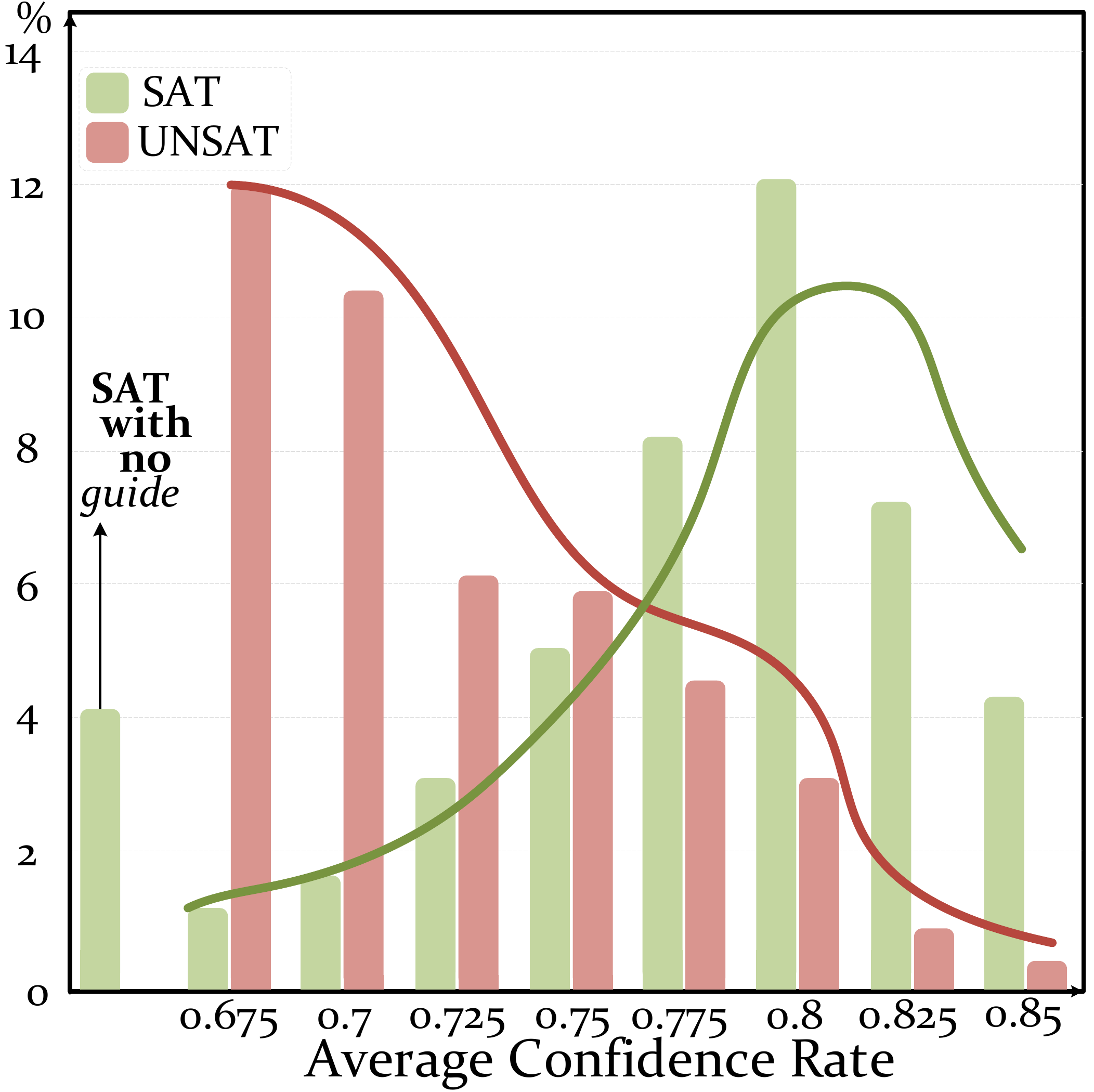}}} \\ \vspace{-10pt}
    \subfloat[]{{\includegraphics[width=0.5\columnwidth]{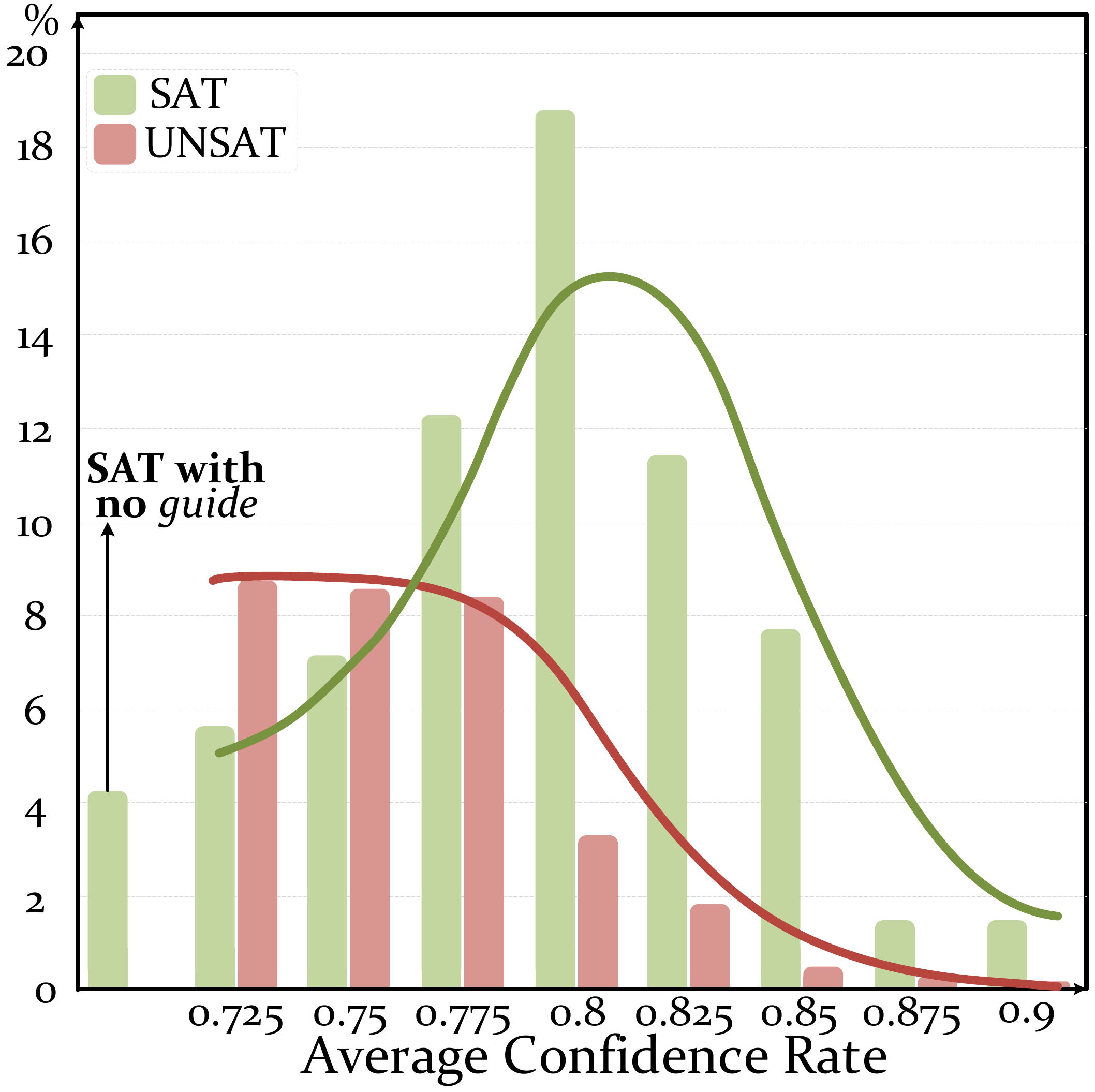}}}
    \subfloat[]{{\includegraphics[width=0.5\columnwidth]{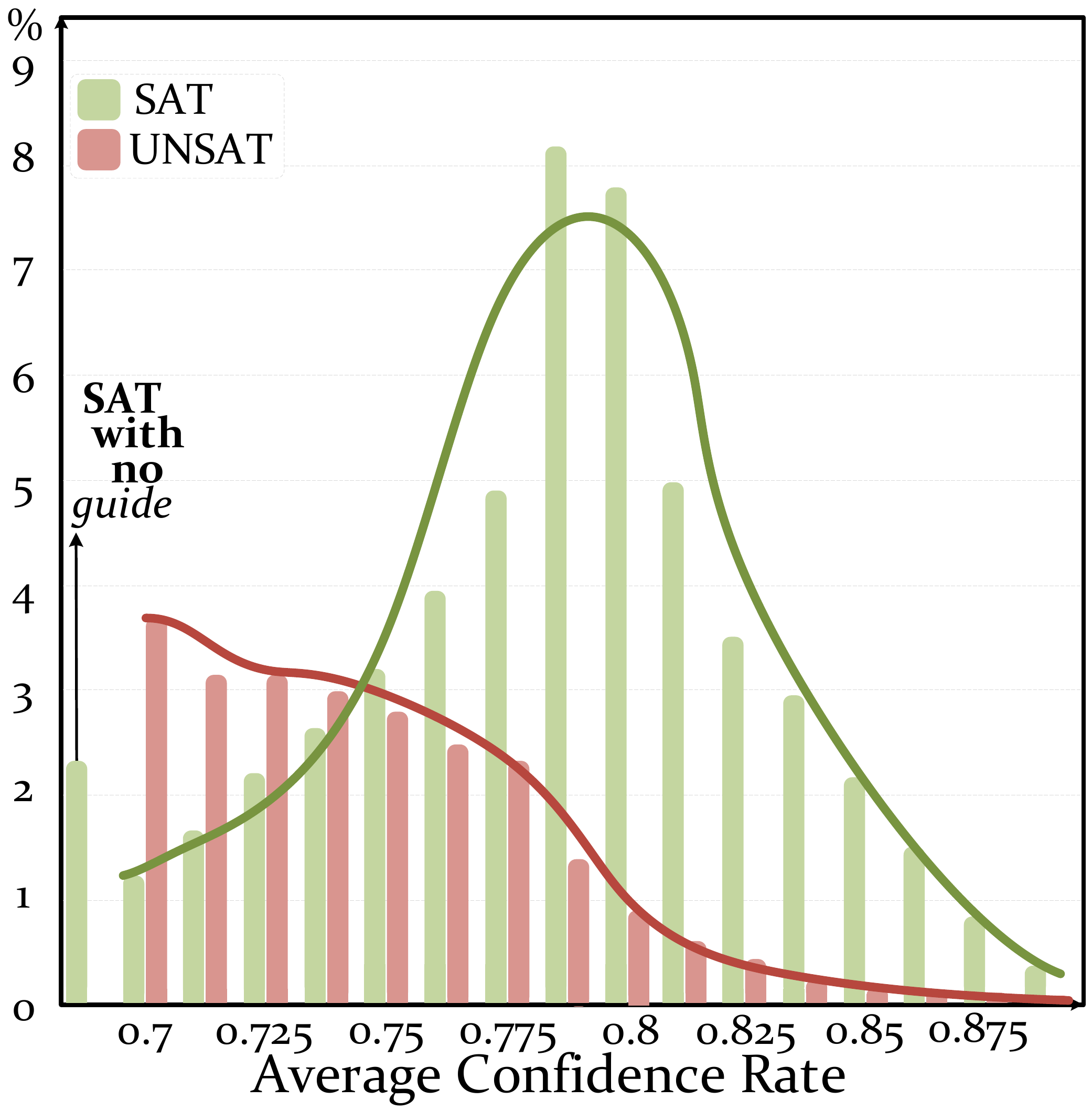}}} \\ 
    \vspace{-10pt}
    \caption{The Impact of Confidence Rate on the Success of MPNN predictions.}
    \label{normal_conf}
\end{figure}

In some cases, the execution time of the traditional SAT attack and that of NNgSAT is equal. This shows that in a non-negligible part of SAT iterations, the network might not provide useful guidance for the actual SAT attack. Hence, when the execution time is identical, it means that none of the network's predictions was helpful for the SAT attack for that case and within $N$ SAT iterations. Also, since the neural network guidance is prediction-oriented, due to misguiding that will happen in a few cases, regardless of the size of the circuit, the performance of NNgSAT might be varied.  

This observation shows that the predictions could be categorized based on their effectiveness: (1) \emph{guiding prediction} that correctly guides its corresponding SAT solver to find a SAT assignment, (2) \emph{misguiding prediction} that incorrectly guides its SAT solver to an \emph{UNSAT}, (3) \emph{skipped prediction} that are skipped/terminated because one of the parallel SAT solvers on this problem found a solution. The ratio of each category is dependent on the CR we choose for each run. Fig. \ref{normal_conf} shows the distribution of two first categories\footnote{Due to the parallelism, there are some SAT solvers per each iteration, in which the only one might find the SAT assignment and all others must be skipped. Hence, the ratio of skipped prediction is far higher and omitted from Fig. \ref{normal_conf}} based on different values of CR. As shown in Fig. \ref{normal_conf}(a, b), when we start the prediction with lower average CR, the rate of misguiding prediction is considerably higher. The predictions help more when the average confidence ratio is 0.7<$CR$<0.9. Hence, as shown in \ref{normal_conf}(c, d), to get the most benefit of the network, CR must be 0.7<$CR$<0.9.

\begin{figure}[t]
    \centering
    \includegraphics[width=\columnwidth]{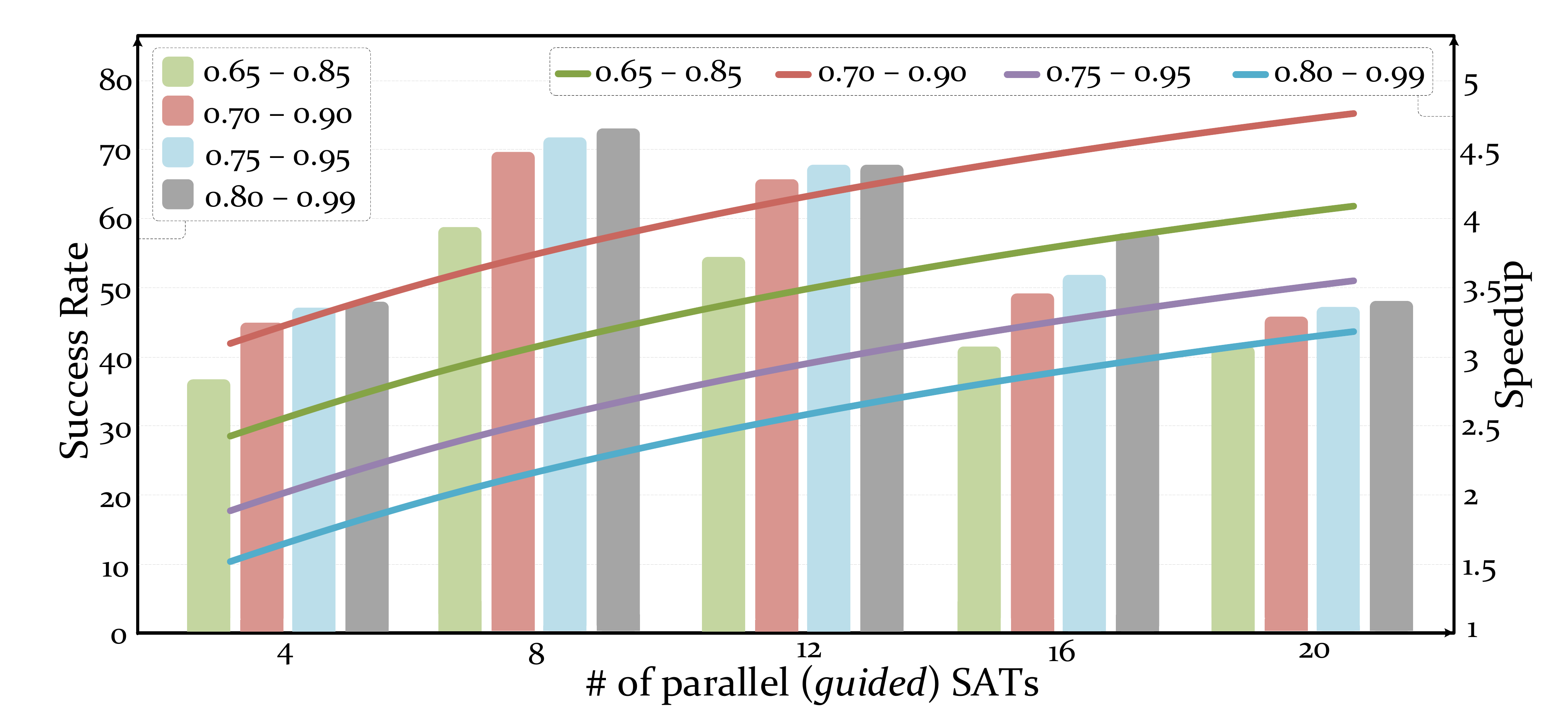}
    \vspace{-15pt}
    \caption{The Impact of Number of Parallel SAT instances on Success Rate and Speedup of NNgSAT. \vspace{-5pt}}
    \label{parallel}
\end{figure}

As we discussed previously, based on the CR, a different number of parallel SAT solvers guided by the MPNN would be executed. To examine the impact of the number of parallel SAT executions on the performance of NNgSAT, we compare the NNgSAT runtime for different numbers of parallel executions over different CR ranges. Fig. \ref{parallel} shows that when we set the CR on a higher range (e.g. 0.8-0.99), as discussed previously, it requires more message passing to guide the SAT. Hence, the performance of the higher ranges will be decreased. However, when we use more cores, it could help to improve the speed-up. On the other hand, for higher ranges, the success rate is not considerably higher. For CR 0.8-0.99, the success rate is only 5\% higher than that of CR 0.7-0.9. However, since for 0.7<CR<0.9, it is less restricted, the performance is significantly higher\footnote{For CR 0.65-0.85, although it is less restricted than CR 0.7-0.9, the performance is lower because in many cases, the prediction is misguided leading to UNSAT.}\footnote{All the results and the speedup reflected in Fig. \ref{parallel} are for those cases in which both attacks, the traditional SAT and NNgSAT, successfully de-obfuscate the locked circuits. This cannot be used as the total speedup of NNgSAT vs. the traditional SAT because for many cases NNgSAT solves problems that were not solvable by the traditional SAT, and in these cases, the speedup cannot be calculated.}. Hence, the number of cores is set to 8, and CR is 0.7-0.9 to have the best tuning for the NNgSAT. Based on these observations, to gather results for Table \ref{sat_exe}, we used 8 cores (8 different CRs) and the CR range is set to 0.7-0.9.

\section{Conclusion} \label{conclusion}

Although the SAT attack (or one of its derivatives) successfully de-obfuscates many state-of-the-art logic locking techniques, there exist families of complex \emph{hard-to-be-solved} structures that remain challenging for the conventional SAT solver. In this paper, we proposed a neural-network-guided SAT attack, called NNgSAT attack, which uses a Message Passing Neural Network to predict satisfying assignments that could help and significantly speedup the conventional SAT attack in solving such cases. Our experimental results show that NNgSAT could de-obfuscate 93.5\% of the logic circuits locked with or contained complex structures within a reasonable time, while the original SAT attack fails to de-obfuscate them.

\section*{Acknowledgement} \label{Acknowledgement}

This research is supported by National Science Foundation (NSF, \#1718434).

\bibliographystyle{ACM-Reference-Format}
\bibliography{refs}

\end{document}